\documentclass[aps,preprint]{revtex4}
\usepackage{amsmath}
\usepackage{graphics}
\usepackage{graphicx}
\usepackage{color}

\usepackage{epstopdf}
\usepackage{here,epsfig,amsfonts,latexsym,color,wrapfig}

\usepackage{amssymb}

\newcommand{\be}{\begin{equation}}
\newcommand{\ee}{\end{equation}}
\newcommand{\bc}{\begin{center}}
\newcommand{\ec}{\end{center}}
\newcommand{\bq}{\begin{quote}}
\newcommand{\eq}{\end{quote}}
\newcommand{\bmat}{\begin{pmatrix}}
\newcommand{\emat}{\end{pmatrix}}

\newcommand{\br}{\mathbf{r}}

\newcommand{\ber}{\mathbf{e}_r}
\newcommand{\bet}{\mathbf{e}_\theta}
\newcommand{\bex}{\mathbf{e}_x}
\newcommand{\bey}{\mathbf{e}_y}
\newcommand{\bez}{\mathbf{e}_z}
\newcommand{\maxx}{\text{max}}
\newcommand{\bB}{\mathbf{B}}

\newcommand{\mpp}{{M}_{PP}}
\newcommand{\mpt}{{M}_{PT}}
\newcommand{\mtp}{{M}_{TP}}
\newcommand{\Mtt}{{M}_{TT}}

\newcommand{\Tdemi}{$T\frac{1}{2}$ flow}
\newcommand{\Sz}{${\cal S}_z$}

\newcommand{\bJ}{\mathbf{J}}

\newcommand{\bnabla}{\mathbf{\nabla}}
\newcommand{\disp}{\displaystyle}
\newcommand{\Rm}{{\rm R}_{\rm m}}

\newcommand{\bv}{\mathbf{v}}

\newcommand{\bV}{\mathbf{V}}

\newcommand{\bn}{\mathbf{n}}
\newcommand{\bemf}{\mathbf{e}}

\newcommand{\cL}{{\cal L}}

\begin{document} 

\title{Dynamo action in cylindrical annulus}

\author{R. Volk, P. Odier, J.-F. Pinton}
\affiliation{Laboratoire de Physique, CNRS UMR\#5672, \\
\'Ecole Normale Sup\'erieure de Lyon, 46 all\'ee d'Italie, F-69007 
Lyon (France)\\
{ } \\
\today \\
{ } \\
{ }
}

\begin{abstract}

We study numerically the induction mechanisms generated from an array of helical motions distributed along a cylinder. Our flow is a very idealized geometry of the columnar structure that has been proposed for the convective motion inside the Earth core. Using an analytically prescribed flow, we apply a recently introduced iterative numerical scheme~\cite{BourgoinPOF} to solve the induction equation and analyze the flow response to externally applied fields with simple geometries (azimuthal, radial, etc.). Symmetry properties allow us to build selected induction modes whose interactions lead to dynamo mechanisms. Using an induction operator formalism, we show how dipole and quadrupole dynamos can be envisioned from such motions. 

\end{abstract}

\maketitle

\newpage

\section{Introduction}
The understanding of the self-sustained dynamo of the Earth is still a major challenge. Following Larmor's original hypothesis~\cite{Larmor}, it is supposed to originate in the convective motion inside the liquid iron core. There, the induction due to fluid motions may overcome the Joule dissipation and a dynamo can be generated. Although knowledge of the fluid motion is essential because it drives the magnetic induction, the structure of the core convective motion is not precisely known. It is due to the extreme range of parameters in this problem: the Earth is in rapid rotation, and the electrical conductivity of molten iron is large, but finite. As a result, two important dimensionless parameters of the problem are very small: the Eckman number may be as low as $10^{-15}$ and magnetic Prandtl number of the order of $10^{-6}$~\cite{Poirier} are usually quoted.  In addition, the source of the convective motion is mixed and still debated: thermal and compositional convection contribute, with heat exchanged at the boundaries (solidification of the iron at the solid inner core, transfer at the mantle boundary) and heat is also released in the bulk because of radioactivity~\cite{Labrosse}. The Rayleigh and Nusselt numbers are not precisely known, and the coupling with the magnetic field can result in significant changes in the structure of the convective flow. However, many models of the Earth dynamo have invoked the columnar flow structure first derived by F. Busse~\cite{Busse} at onset of convection in a spherical Couette geometry~\cite{DynamoColumns}. 
These columns may persist for a system moderately above the onset of convection~\cite{Dormy,Morin}  although their existence in the actual Earth core is still an open question~\cite{Harder}. There are also arguments for a subcritical dynamo bifurcation~\cite{Subcritical}, in which the columnar structure may be stabilized by the large scale magnetic field~\cite{Busse2002}. 

The focus of this paper is on the induction mechanisms that result from the existence of such a columnar structure. We consider a highly idealized situation in which columns with helical motion are distributed along a cylinder, as helicity is known to favor dynamo action~\cite{Moffat}. The geometry is chosen to be cylindrical so that the conical and curves conditions of the Earth are ignored. We also do not take into account the drift of Busse's rolls, nor the presence of a zonal wind. 
The flow is prescribed, and we analyze its response to externally applied fields with simple geometries (azimuthal, radial, etc.) in order to  understand the induction mechanisms in this system. For instance, several numerical simulations (e.g.~\cite{Grote} and references therein) have noticed that both dipole and quadrupole dynamos are possible, with thresholds in close proximity. Note that the periodic array of helical columns in the Roberts flow~\cite{Roberts} and Karlsruhe dynamo~\cite{Karlsruhe} cannot generate an axial dipole. We show here that the annular geometry can lead to both axial dipoles and quadrupoles. We describe in detail the flow structure and the numerical method in section II. In sections III  we present our results concerning induction responses to simple applied fields. Section IV is devoted to the study of possible dynamo action in this type of flows, based on an operator formalization. Conclusions and possible extension of the study are presented in section V.


\section{System and methods }

\subsection{System}
The flow geometry is made of a system of cylindrical columns forming an annulus, as shown in figure~1. The columns are grouped in pairs of cyclonic and anti-cyclonic roll, for which the axial flow is reversed. The velocity is assumed to be stationary and is expressed as the sum of 
a contribution due to the circular motion of the fluid in a column (the {\it rotational} component $\bV^R(\br)$) and of a contribution due to an {\it axial} motion, which we label $\bV^A(\br)$. In an Earth-like geometry, $\bV^A(\br)$ would be generated by the Eckman pumping at upper and lower boundaries. One thus writes

\be\label{equ:velocity}
\bV(\br) = \bV^R(\br) + \xi \bV^A(\br) \ ,
\ee

We chose analytical expressions of the  fields such that the components are separately divergence-free, $\bnabla \cdot \bV^R(\br) = \bnabla \cdot \bV^A(\br) = 0$. The coefficient $\xi$ measures the intensity of the axial motion compared to the rotational one. 

The rotational part in columns of height $2H$ is expressed in cylindrical polar coordinates as:

\be
\bV^R \left \{
\begin{array}{l}
\disp
  V^{R}_r(r,\theta) = \sin(n \theta) \cdot \sin\left[\frac{\pi}{d} \left( r-(R-d) \right) \right]\\
\disp
  V^{R}_\theta(r,\theta) = \frac{1}{n} \cos(n \theta) \cdot \left\{ \frac{\pi r}{d} \cos\left[ \frac{\pi}{d} \left( r-(R-d) \right)  \right] + \sin \left[ \frac{\pi}{d} \left( r-(R-d) \right) \right] \right\}\\
\disp
  V^R_z(r,\theta) = 0 \\
\end{array}
\right.
\ee
where $n$ is the number of column pairs, $R$ is the outer radius of the annulus, $d$ is the thickness of the region in which the columns are confined (within radial distances between $R-d$ and $R$). The velocity is set to zero outside the domain $R-d  \leq r \leq R$. 

For the axial flow, we consider two cases. In the first one, the columns have a height $2H$ and the axial flow has a defined direction within each column and reverses between neighboring columns. It corresponds to the geometry sketched in figure~\ref{fig:geometry}(a). Figure ~\ref{fig:geometry}(c) and (d)  show respectively a cut of the flow in the planes $z=0$ and $\theta=0$. The helicity ($\cal{H}=\mathbf{V}\cdot(\nabla\times\mathbf{V})$) has the same sign in each column, negative in this case (a column with positive axial velocity rotates with negative vorticity). The corresponding poloidal velocity is given by:

\be\label{equ:Tdemiflow}
\bV^A \left \{
\begin{array}{l}
\disp
  V^{A}_r(z,\theta) = 0 \\
\disp
  V^{A}_\theta(z,\theta) = \frac{\pi r}{2nH}  \sin (n \theta) \sin \left( \frac{\pi z}{2H} \right) \\
\disp
  V^A_z(z,\theta) = \cos(n\theta) \cos \left( \frac{\pi z}{2H} \right) \\
\end{array}
\right.
\ee
In the second case (sketched in figure~\ref{fig:geometry}(b)), the axial flow is reversed about the plane $z=0$, hence having symmetries similar to Busse's columns in rotating convection. Figure ~\ref{fig:geometry}(e) shows a cut of the flow in the plane $\theta=0$. The helicity is therefore negative in the columns of the upper half of the cylinder and positive in the lower half. The axial velocity is then written:

\be\label{equ:T1flow}
\bV^A \left \{
\begin{array}{l}
\disp
  V^{A}_r(z,\theta) = 0 \\
\disp
  V^{A}_\theta(z,\theta) = - \frac{\pi r}{nH}  \sin (n \theta) \cos  \left( \frac{\pi z}{H} \right)\\
\disp
  V^A_z(z,\theta) = \cos(n\theta) \sin  \left( \frac{\pi z}{H} \right) \\
\end{array}
\right.
\ee

In our study, we call $T\frac{1}{2}$ the configuration obtained with the rotational velocity and the first choice of the axial velocity field, and $T1$ the configuration obtained with the second choice. We shall look for stationary solutions  of the induction equation: 
\be
\partial_t \bB = 0 = \bnabla \times ( \bV \times \bB) + \eta \Delta \bB \ ,
\ee

where $\eta$ is the magnetic diffusivity of the fluid. The boundary conditions are such that the medium inside the annulus has the same electrical conductivity as the fluid while the outside medium is insulating. The magnetic permeability is equal to that of vacuum in the whole space.

We stress that once the radius $R$ of the cylinder  and the aspect ratio $H/R=2$ are fixed there are still many independent parameters which may be varied: the number $2n$ of columns, their aspect ratio $d/R$, the magnitude of the axial flow compared to the rotational one, etc. We shall concentrate on the geometry portrayed in figure~1: four pairs of columns with relative thickness equal to 0.4 (yielding a square aspect ratio for the columns), and $\xi=1.25$ (this sets the ratio of the maxima of axial to rotational velocities to 0.7, close to the values in the existing experimental dynamos~\cite{Karlsruhe, Riga, vksdynamo}). The only remaining free parameter is the amplitude of the velocity field, which is non-dimensionalized in the form of the magnetic Reynolds number $\Rm= V_{\rm max} R / \eta$.  $V_{\rm max}$ is the maximum velocity in the domain of the flow, and we write $\bV = V_{\rm max} \bv$.

\subsection{Iterative procedure}
We use here the iterative technique introduced in~\cite{BourgoinPOF,stefani}.  The reader is referred to it for a detailed presentation, and for an evaluation of its performance compared to standard analysis in magnetohydrodynamics, such as the linear stability analysis of the induction equation. We stress that the technique is best suited for the study of stationary solutions, and only recall here its basic principles. 

We consider the response of the flow to an applied magnetic field $\bB_0$, looking for the induced field $\bB$ which solves the stationary induction equation
\be
\bnabla \times ( \bV \times \bB) + \eta \Delta \bB = - \bnabla \times ( \bV \times \bB_0) \ .
\ee
The result is expressed as the integer series 
\be
\bB = \sum_{k=1}^{\infty} \bB_k \; \; \; {\rm with} \; \; \; |\bB_k| \sim {\cal O}(\Rm^k) B_0 \ ,
\ee
The contributions $\bB_k$ are computed iteratively  from a hierarchy of non-dimensional Poisson equations
\be\label{equ:poisson}
\Delta \bB_{k+1} = - \Rm \bnabla \times (\bv \times \bB_k ) \ ,\;\;\;\; k\ge 0
\ee
which can be solved for any given set of boundary conditions.

\subsection{Numerical simulation}
Applying a standard Poisson solver to equation~(\ref{equ:poisson}) would require to write the complete set of conditions for the magnetic field at the boundaries of the computational domain. In practice, the boundary conditions are more readily expressed in terms of electric currents and potentials. We thus implement the following sequence:

\noindent (i) The electromotive force (e.m.f. in units of $V_{\rm max} B_0$) induced by the flow motion is  computed from $\bemf_{\rm k+1} = \bv \times \bB_k$.

\noindent (ii) Electric current being divergence free, the distribution of electric potential is obtained from $\Delta \phi_{k+1} = \bnabla \cdot (\bv \times \bB_k )$, with von Neuman boundary conditions ($\bn \cdot \bnabla \phi = \bn \cdot (\bv \times \bB)$, with $\bn$ the outgoing normal of the domain).

\noindent (iii) Induced currents (non-dimensionalized) are then computed as from Ohm's law $\mathbf{j}_{k+1} = -\bnabla\phi_{k+1}+\bemf_{k+1}$, and used to compute the magnetic field $\bB_{k+1}$ from Biot and Savart law. \\

All calculations are made with a magnetic Reynolds number equal to unity (we set $R=1$, $V_{max}=1$ and $\eta=1$). Its actual value enters only in the final step, when contributions are collected and the integer series is computed: $\bB(\Rm) = \sum_k \bB_k(\Rm=1) \Rm^k$. This approach requires that the series converges, and sets an upper value for the magnetic Reynolds number $\Rm$ (${\Rm} < \Rm^\ast$).  We have found that the radius of convergence ${\Rm}^\ast$ is of the order of 30; for higher magnetic Reynolds number values we have shown in~\cite{BourgoinPOF} that Pad\'e approximants~\cite{Pade} still give results in remarkable agreement with the solution of the induction equation computed without approximation.


\section{Induction processes.}\label{Tdemi}

\subsection{Induction from a toroidal applied field.}\label{sec:orthoradial}
When an external magnetic field is applied in the azimuthal direction ($\bB_0 = B_0\bet$), one expects the generation of an azimuthal current, in very much the same manner as the $\alpha$ effect~\cite{Radler} operates in the Roberts flow~\cite{Roberts} and in the Karlsruhe dynamo~\cite{Tilgner,Karlsruhe}. This is because there is no conceptual difference between a horizontal field transverse to the columns in the Karlsruhe geometry and a toroidal field applied in the geometry considered here. In the azimuthal direction, there is a scale separation between the size of one column and the circumference of the annulus, so the results of mean-field MHD, as for instance explored in~\cite{Raul}, should apply. 

\subsubsection{Induction at first order}
We start with induction at  first-order, i.e. the solution of the first iterative step of the induction equation:
\be\label{induc}
\Delta \bB_{1} = - \bnabla \times (\bv \times \bB_0 )
\ee 

The topology of $\bB_0$ is chosen identical to the one that would be generated by the flow of an electric current in a rod of radius $a$ located along the $z$-axis
\be
\bB_0 =  B_0 \left \{
\begin{array}{l}
\disp
 \frac{r}{a} \bemf_\theta \; \; \; (r \leq a=0.1)\\
\disp
 \frac{a}{r} \bemf_\theta \; \; \; (a \leq r)\\
\end{array}
\right.
\ee
The induced field $\bB_1$ is shown in figure \ref{fig:BthetaB1}(a) and (b). One observes two major effects: in the axial direction the field lines are stretched by the vertical velocity gradients, and in the transverse plane they are advected by the rotational motion in the columns. The mechanisms are schematically shown in figure~\ref{fig:BthetaB1}(c) and (d); $B_{1,z}$ (figure~\ref{fig:BthetaB1}(c))  results from the action of the axial flow, as can be seen by writing the axial projection of equation (\ref{induc}):
\be
\Delta B_{1,z} = - B_0 \frac{a}{r} \partial_\theta v^A_{z} \ .
\ee
Note that since the stretching is largest at the boundary between two columns, the induced field is also maximum at these points, therefore out of phase by a quarter of a period with respect to the velocity field. 

For the transverse components of the induced field, $(r, \theta)$ directions, all azimuthal gradients of the velocity contribute {\it a priori}
\be
(\Delta \bB_1)_{(r,\theta)} = B_0 \frac{a}{r^2} \partial_\theta \bv^R_{(r,\theta)} + \bv^R\cdot\nabla\bB_0 - B_0 \frac{a}{r^2} \partial_\theta \bv^A_{(r,\theta)} +  \frac{1}{r}\bv^A_\theta \cdot\partial_\theta \bB_0 \ .
\ee
However, the picture is simpler when restricted to induction in the $z=0$ plane. In this case the last two terms of the above equation do not contribute because $\bv^A$ is an odd function with respect to $z$. In particular, one obtains for the field induced in the radial direction
\be 
\Delta B_{1(r)} = B_0 \frac{a}{r^2} \left( \partial_\theta v^R_r - v^R_\theta \right) + \frac{v^R_\theta}{r} B_0 \frac{a}{r} =  B_0 \frac{a}{r^2}  \partial_\theta v^R_r \ ,
\ee
which shows that (i) the stretching by the azimuthal gradient of $v^R_\theta$ is compensated by the advection of the applied field; (ii) the net effect is due to the azimuthal modulation of the radial flow (figure~\ref{fig:BthetaB1}(d)). At first order, the rotational flow component generates in the center of the columns an induced field perpendicular to the applied one, a mechanism which is quite helpful in the understanding of field expulsion by vortices. 

\subsubsection{Second order effects}
We now turn to the analysis of the induction at second order of iteration, {\it i.e.} the induction resulting from the presence of $\bB_1$. The electromotive force $\mathbf{e}_2 = \bv \times \bB_1$ is shown in figure~\ref{fig:BthetaB2}(a,b). At this stage, the induction from each column add-up cooperatively. An azimuthal current $J_{2,\theta}$ is induced -- figure~\ref{fig:BthetaB2}(b) -- parallel to the applied field $\bB_0$. It has the symmetry of the flow helicity. As we have verified, it is reversed as either $\bv^A$ or $\bv^R$ is reversed but not when both change sign. $J_{2,\theta}$ generates the axial field $B_{2,z}$ shown in figure~\ref{fig:BthetaB2}(c). This induction by helical motion, schematized in figure~\ref{fig:BthetaB2}(e), was first proposed by Parker~\cite{Parker}, and evidenced in the VKS experiment~\cite{vksalpha}. It is also in agreement with the $\alpha$ effect introduced in the framework of mean-field magnetohydrodynamics~\cite{Radler}, for the 2D array of columns in the Roberts flow~\cite{Roberts}, and for the columnar ring discussed here~\cite{Raul}. Following this analogy, we will call this effect $\alpha$-effect in the rest of our study.

The electromotive force ${\mathbf e}_2$ also generates axial currents which create the azimuthal field $B_{2,\theta}$ shown in figure~\ref{fig:BthetaB2}(d): it is largest in the center of the columns and its direction is opposed to the applied field. Its generation -- schematized in ~\ref{fig:BthetaB2}(f) -- shows that $B_{2,\theta}$ is opposed to the applied field even when the rotation of the columns is reversed, or when the axial flow is reversed or even suppressed. The effect traces back to the expulsion of magnetic field lines by vortical motions, as shown by numerical~\cite{Weiss} and experimental~\cite{Odier2000} studies on an isolated vortex. It is a second order effect: as pointed out previously, the rotational motion of the column can be thought of as turning the applied field lines by 90 degrees at each step, so that after two iterations the induced field is directly opposite to the applied one. 

We stress that the response of the flow to an applied field is more complex than the generation of an azimuthal current $J_{2,\theta}$ by an `$\alpha$' effect due to the helical motion in the columns; this effect is indeed present, but the net current ${\mathbf J}_2$ is related to the applied field by a full second rank tensor, ${\mathbf J}_2 = \sigma [\alpha] \bB_0$. In the tensor $[\alpha]$ not all components are due to the flow helicity and some components correspond to an expulsion effect. 

\subsubsection{Azimuthal averages}
In the remaining of our study, we consider the contributions to the induction averaged over the azimuthal angle $\theta$, as they present a somewhat clearer picture. Figure~\ref{fig:BthetaB2moy} shows the averaged order 2 field, defined as:
\be
\langle \bB_2 \rangle (r,z)  \equiv \frac{1}{2\pi} \int_0^{2\pi}  d\theta \, \bB_2(r,\theta,z) \, 
\ee
One clearly sees the induced azimuthal current $\langle J_{2,\theta}\rangle$ and the associated magnetic field with a dipolar structure. In a finite geometry with insulating surroundings, the currents $\langle{\mathbf J}_2\rangle$ must remain within the flow volume, leading to poloidal current loops (arrows in \ref{fig:BthetaB2moy}(a)) which participate to the expulsion of the applied field. 

In order to quantify the importance of each effect, we have plotted in figure~\ref{fig:BthetaB2prof} the evolution of the components of $\langle \bB_2 \rangle$ with the radial and axial positions. We observe in figure~\ref{fig:BthetaB2prof}(a) that the dipolar field $\langle B_{2,z}\rangle$ is about twice as large as the component $\langle B_{2,\theta}\rangle$ of field expulsion. In figure~\ref{fig:BthetaB2prof}(b), the induction is maximum in the mid-plane $z=0$, save for the $\langle B_{2,r}\rangle$ component for which the top and bottom boundary conditions play a major role.

\subsubsection{Higher orders}\label{def_L}
As detailed in~\cite{BourgoinPOF}, the interest of the iterative procedure is to associate induction effects with specific actions of velocity gradients. It is particularly convenient when a patterns develops through the iterations. As it is the case here, we define a tool to help us quantify the convergence of the pattern. Let $\bB_j$ and $\bB_k$ be the fields induced at respective orders $j$ and $k$. We compute the scalar product
\be\label{equ:scalarprod}
( \bB_j | \bB_k ) = \frac{1}{2 \pi R^2 H}\int_V d^3r \; \bB_j \cdot \bB_k  \ ,
\ee
and the associated norm ${\cal N}(\bB_k)=\sqrt{( \bB_k | \bB_k )}$. Comparisons are made using the normalized scalar product
\be
P_{j,k} = \left( \frac{\bB_j}{{\cal N}(\bB_j)} \left|  \frac{\bB_k}{{\cal N}(\bB_k)} \right. \right) \ .
\ee

In the case of an azimuthal applied field, the expulsion eventually dominates. To wit, we compare in figure~\ref{fig:BthetaBSup} the induced fields at order 1 and 3, and successive iterations of the axial field $\langle \bB_{2k} \rangle$. One finds $P_{1,3} \sim -0.9$, while $P_{2,3} \sim 0$: the field induced at third order is almost exactly opposed to $\bB_1$ and perpendicular to $\bB_2$. As a result, the successive induction steps lead to the expulsion of the applied field, with $P_{k, k+1} \sim 0$ and $P_{k, k+2} \sim -1$, at higher orders. Bear in mind that this concerns normalized values. In dimensional units, one has :

\be\label{equ:loop_orthoradial}
 \bB_{k+2} \sim \gamma  \bB_k \;\;\;\;,{\rm with}\;\; \gamma \sim -1/400
 \ee
 
After the 10$^{\rm th}$ order, we could not detect any appreciable evolution of the pattern. 

\subsubsection{Evolution with $\Rm$}
The magnetic Reynolds number is re-introduced in the summation $\bB_{\rm ind} = \sum_{k=1}^{\infty} \Rm^k \bB_k$. We have used the summed terms up to order 22, resulting in a 1\% accuracy. We show in figure~\ref{fig:BthetaBSupProfiles} the evolution with $\Rm$ of two components of the induction: the axial field in the center of the cylinder, and the toroidal field in the center of the columns. For $\Rm \leq 8$, the calculation at second order yields a very good approximation of the net induction. This is interesting because the 2$^{\rm nd}$ order truncation corresponds to the computation of the mean-field theory with a first order smoothing approximation~\cite{Raul}. For magnetic Reynolds numbers greater than about 8,  the contributions of higher orders in the summation need to be taken into account. One finds that the mean-field approximation tends to over-estimate the induced dipole field -- figure~\ref{fig:BthetaBSupProfiles}(a), as well as the expulsion of the toroidal field -- figure~\ref{fig:BthetaBSupProfiles}(b). 

Empirically, we observe that the radius of convergence of the integer series is $\Rm^\ast = 17$. This value can be understood from our observation that $P_{k,k+1} \sim 0$ and $\bB_{k+2} \sim \gamma \bB_k$. Indeed one can then rewrite the summation as
\be
\bB = (\Rm \bB_1 + \Rm^2 \bB_2) \sum_{k=0}^{\infty} (-|\gamma| \Rm^2)^k \ ,
\ee
from which one immediately gets $1/\sqrt{| \gamma |} \sim 20$ for the radius of convergence. In addition, for $\Rm<\Rm^\ast$, one gets:
\be
\bB \simeq \frac{\Rm}{1 + | \gamma | \Rm^2} \bB_1 +  \frac{\Rm^2}{1 + | \gamma | \Rm^2} \bB_2 \ ,
\ee
which shows that the divergence of the integer series actually lies in the existence of imaginary roots. Such a configurations is particularly suited to the use of Pad\'e approximants~\cite{Pade}. The result, plotted as a solid line in figure~\ref{fig:BthetaBSupProfiles}, shows that at large magnetic Reynolds numbers ($\Rm\ge40$) the axial induction and the expulsion may saturate.

\subsection{Induction from a radial field applied}\label{sec:radial}
As detailed above, starting from an applied toroidal field, the $\alpha$-effect generates a poloidal induced field with a large axial component, but it is easy to show that this axial component gives in turn very weak contributions to the induction. Indeed, an applied field in the axial direction induces fields that are two orders of magnitude weaker than the values obtained with other orientations. In contrast, we show in this section that a radially applied field generates an induced field which has a significant component in the azimuthal direction. 

Specifically, we consider an applied field $\bB_0$ of the form
\be
\bB_0 =  B_0 \left \{
\begin{array}{l}
\disp
 \frac{r}{a} \bemf_r \; \; \; (r \leq a=0.1)\\
\disp
 \frac{a}{r} \bemf_r \; \; \; (a \leq r)\\
\end{array}
\right.
\ee
Two remarks about this functional form are in order: 

(i) within the domain of resolution of the induction equation it is essential that $\bB_0$ be divergence free because we solve iteratively $\Delta \bB_{k+1} = - \nabla \times (\bv \times \bB_k)$ rather than $\Delta \bB_{k+1} = -\bv \cdot \nabla \bB_k - \bB_k \cdot \nabla \bv$.

(ii) here, $\bB_0$ is not divergence free for $r \leq a$, but in this domain the source term in $\Delta \bB_1 = -\nabla \times (\bv \times \bB_0)$ vanishes with $\bv$. 

\subsubsection{Induction at first order}
The induced field at first order is shown in figure~\ref{fig:BrB1}. We observe in (a) and (b) that the magnetic induction is in phase with the flow in the columns, as opposed to the case of a toroidal applied field, where there was a phase difference of a quarter of a period. In the transverse plane, the effect of the azimuthal part of the velocity field is to twist the applied field lines, thereby generating azimuthal components in the magnetic field, in direct correspondence with the case of an azimuthal applied field, shown in figure~\ref{fig:BthetaB1}(d). The magnetic field induced in the axial direction is due to the radial gradient of the vertical velocity and solves $\Delta B_{1,z} = - B_0 (a/r) \partial_r v_z^A$. It corresponds to the stretching of applied field lines in the vertical direction. However, it is can be seen in figure~\ref{fig:BrB1}(a,c) that the induction lies essentially near the boundary with the inner core ($r \sim 0.6$) compared to the boundary with the outer medium ($r\sim 1$). Since at both location there exists a large radial gradient in the vertical velocity, the difference has to be linked with the electromagnetic boundary conditions. The inner core has the same conductivity as the fluid while the outer medium is insulating, so that the magnetic field is potential outside the cylinder. For a $z$-independent flow, one can show that this condition ensures that the induced field $B_z$ vanishes at the $r=1$ boundary~\cite{CommentBz}.   In the case of our flow, the velocity field is not $z$-independent but its components have only second order variation with $z$ in the mid-plane.  One thus understands why the induction in the axial direction is so weak at $r=1$, as shown in figures~\ref{fig:BrB1}(a,c).

\subsubsection{Induction at second order}\label{radial_order2}
As can be expected, in correspondence to the induction in the case of an azimuthal applied field, the screw motion in the columns produces again a current parallel to the applied field. This is shown in figure~\ref{fig:BrB2}(b) where one finds an induced radial e.m.f. maximum in the center of each column. This electromotive force has the symmetry of the helicity, reversing if either $\bv^A$ or $\bv^P$ are reversed (but not both simultaneously). The corresponding induced currents generate the axial field $B_{2,z}$ shown in figure~\ref{fig:BrB2}(c). The field is periodic in the azimuthal direction with a period equal to half the period of the column distribution -- in agreement with the Parker induction mechanism sketched in figure~\ref{fig:BthetaB2}.e (vertical arrows). 

When one computes the azimuthal average, a first observation is that the currents yield a non zero contribution $\langle \bJ_2 \rangle$ which is essentially radial in the mid-plane $z=0$. In addition the current lines close inside the conducting fluid, leading to the poloidal loops shown in figure~\ref{fig:BrB2}(e). They generate a field $\langle\bB_2\rangle$ whose toroidal part is anti-symmetric about the $z=0$ plane, as shown in figure~\ref{fig:BrB2}(f). 

One thus finds that a toroidal magnetic field is generated from the application of a radial magnetic field. The conversion is less efficient than the reverse process, discussed in the previous section. One computes  ${\rm max}\{{\langle B_{2,\theta}\rangle / B_{0,r} \}} \sim 7 \, \cdot 10^{-4}$ instead of ${\rm max}\{{\langle B_{2,r}\rangle / B_{0,\theta} \}}\sim 11 \, \cdot 10^{-4}$. However, the major benefit is that the expulsion is much weaker. The component of $\langle\bB_2\rangle$ opposed to the applied radial field is 10 times weaker than the induced toroidal field. This can be explained by the fact that, while $\bB_2$ has a component opposed to the applied field in the center of the columns, it has a contribution which reinforces the applied field in their periphery as can be seen in figure~\ref{fig:BrB2}(d). Therefore, when the azimuthal average is taken, the expulsion effect is weakened. On the contrary, in the case of the azimuthal applied field, the expulsion effect from each columns add up collectively (see  figure~\ref{fig:BthetaB2}(b)), resulting in a larger contribution.

Hence, expulsion is less effective in the direction perpendicular to the direction where scale separation develops (in the azimuthal direction the columns cross-section is an order of magnitude smaller than the cylinder diameter, while in the radial direction the characteristic size of the flow is equal to the width of the column). We conclude that in the case considered here, scale separation does not particularly favor magnetic induction, but dramatically reduces expulsion.

\subsubsection{Higher orders}
The structure of the induced magnetic field is rapidly stabilized as higher orders are computed. As shown in figure~\ref{fig:BrBSup} the fields at orders 3 and 5 closely resembles that at order 1; the normalized scalar product are $P_{1,3}  \sim -0.5$ and $P_{3,5} \sim -0.9$.

The fields produced at a given order are again fairly orthogonal to the fields at next or previous order : $(\bB_k | \bB_{k+1}) \sim 0$. In addition, as can be seen in figure~\ref{fig:BrBSup}(g), the iteration  converges for even orders towards a quadrupolar structure with a negative feed-back in a two step mechanism:
\begin{equation}\label{equ:loop_radial}
\bB_{k+2} = -\gamma \bB_k \; \; \; (\gamma \sim 1/415) \ .
\end{equation}

\subsubsection{Evolution with $\Rm$}
The $\Rm$ dependence of the induced radial and azimuthal magnetic field is shown in figure~\ref{fig:BrBSupProf}, after summation of the terms up to order 22. Note that $B_\theta$ is sampled at $(r=0.7, z=-0.6)$, because for a quadrupole, the toroidal field is very small in the $z=0$ plane. One observes in figure~\ref{fig:BrBSupProf}(b) that a second order calculation is a correct approximation for magnetic Reynolds numbers up to 10. For higher $\Rm$ other terms need to be included. As in the case of the toroidal applied field, they tend to slow the increase of the induced field, mainly because of expulsion generated from rotational motion in the columns. The integer series diverges for $\Rm^\ast = 18$, as expected from the value $1/\sqrt{| \gamma |} \sim 20$. Results obtained using the Pad\'e approximants, although extending the computed induction beyond the radius of convergence of the series, do not point to a saturation at large $R_m$.


\section{Dynamo action}

In the previous section, we showed that the induction mechanisms in the case of the $T\frac{1}{2}$ flow consist in a mutual conversion between azimuthal and radial fields, through the $\alpha$ effect, along with an effect of expulsion of these fields by the rotating columns. We have identified two modes, dipole and quadrupole, mainly axisymmetric, that realize a feed-back loop in a two-step mechanism (equations~(\ref{equ:loop_orthoradial}) and (\ref{equ:loop_radial})) with a negative sign, therefore leading to an anti-dynamo configuration. Following the ideas developed in~\cite{BourgoinPOF}, we can express these results using an induction operator formalism: from equation (\ref{equ:poisson}), we define  {$\cL(\Rm) \equiv - \Rm \Delta^{-1} \left\{ \nabla \times (\bv \times \bullet) \right\}$ for a velocity field $\bv$ corresponding to the value $\Rm=1$. Equations~(\ref{equ:loop_orthoradial}) and (\ref{equ:loop_radial}) can then be interpreted in the following way: the $\langle \bB_{2n}\rangle$ modes ($n>5$) obtained in the induction studies in section~\ref{sec:orthoradial} (dipole mode) and \ref{sec:radial} (quadrupole mode)
are eigenvectors of $\cL^2(\Rm)$. From this observation, we will show in this section that using a poloidal/toroidal decomposition, a matrix analysis can be performed on the identified eigenmodes to find eigenvectors of $\cL^2(\Rm)$ with positive eigenvalues, thus leading to possible dynamo solutions.

Indeed, writing for simplicity $\cL=\cL(1)$ (we then have $\cL(\Rm)\bB=\Rm\cL\bB$), let's assume that we can find a magnetic field $\bB_e$ that is not an eigenvector of the operator $\cL$, but of $\cL^2$ with a positive eigenvalue $\gamma_e$. We can easily show that $\bB_s=\bB_e+\frac{1}{\sqrt{\gamma_e}} \cL\bB_e$ is an eigenvector of $\cL$ with a positive eigenvalue $\gamma_s=\sqrt{\gamma_e}$. Taking $\Rm=1/\sqrt{\gamma_e}$, we then have $\cL(\Rm)\bB_s=\bB_s$, which defines $\bB_s$ as a self-sustained magnetic field by the velocity field, at threshold $\Rm=1/\sqrt{\gamma_e}$.

\subsection{Dynamo action in the $T\frac{1}{2}$ flow.}\label{dynamo_Tdemi}

A dynamo cycle is often seen as a toroidal/poloidal feed-back loop~\cite{Moffat}. In the cylindrical geometry of the $T\frac{1}{2}$ flow, it is easy to decompose an axisymmetric field into its poloidal and toroidal part. We perform such a decomposition successively on the dipole and quadrupole axisymmetric eigenmodes of the $\cL^2$ operator and then compute the action of this operator on the vector space generated by these two components. This leads us to a matrix formulation of the problem, allowing the identification of dynamo modes. We then perform the same analysis on a non axisymmetric mode.

\subsubsection{Generation of a dipole field.}\label{sec:dipole}

We start from the field $\langle {\bf B}_{10}\rangle$ -- figure~\ref{fig:BthetaBSup}(e) -- which was shown (equation~(\ref{equ:loop_orthoradial})) to be an eigenvector of $\cL^2$, with a dipolar geometry and a negative eigenvalue. Figure~\ref{fig:L_dipole}(a) and (c) show the decomposition of this field into its toroidal -- $\bB_d^T$ -- and poloidal -- $\bB_d^P$ -- components. 

Figure~\ref{fig:L_dipole}(b) and (d)  show the axisymmetric part of the fields  $\cL^2\bB_d^T$ and $\cL^2\bB_d^P$. As expected, the resulting fields are of opposite sign with respect to  $\bB_d^T$ and $\bB_d^P$. In addition, their topology are very similar and look like a linear combination of the initial fields  $\bB_d^T$ and $\bB_d^P$. Using the scalar products defined in equation (\ref{equ:scalarprod}), we can project $\langle\cL^2\bB_d^T\rangle$ and $\langle\cL^2\bB_d^P\rangle$ onto the initial fields, thus  defining an induction matrix:

\be \label{equ:matrixM}
 M_d(T\mbox{\footnotesize{\it $\frac{1}{2}$}})=\bmat \disp\mpp&\disp\mpt\\
\disp\mtp&\disp\Mtt \emat= \bmat (\langle \cL^2 B^P_d\rangle \mid B^P_d )&(\langle\cL^2 B^P_d\rangle \mid B^T_d)\\
(\langle\cL^2 B^T_d\rangle \mid B^P_d)&(\langle\cL^2 B^T_d\rangle \mid B^T_d) \emat 
\;\;\;,\ee

where we have taken ${\cal N}(\bB_d^T)={\cal N}(\bB_d^P)=1$, using the norm $\cal N$ defined in section \ref{def_L}.

This matrix $M_d(T\mbox{\footnotesize{\it $\frac{1}{2}$}})$ is the restriction of the two-step induction operator $\cL^2$ to the vector space of the axisymmetric dipoles. The diagonal terms represent the expulsion effect and the extra-diagonal terms represent the action of the $\alpha$ effect. 

In the case of the $T\frac{1}{2}$ flow with ($d/r=0.4$, $\xi=1.25$ and $n=4$), we compute:
\be 
M_d(T\mbox{\footnotesize{\it $\frac{1}{2}$}}) = -10^{-4} \bmat 10&21\\8&14 \emat 
\ee

This matrix has a negative eigenvalue $\lambda_d^-=-25.10^{-4}=-1/400$, corresponding to the values obtained by iterating the induction operator in the case of an azimuthal or radial applied field. In addition, it has a positive eigenvalue $\lambda_d^+=10^{-4}$, showing that the flow can actually sustain an axial dipole dynamo, for $\Rm\ge\Rm^c=1/\sqrt{\lambda_d^+}=100$. The eigenmodes corresponding to both eigenvalues have the following structure:

\be
\begin{array}{c}
\bB_d^{+} = 0.88 \bB^P_d - 0.47 \bB^T_d\\
\\
\bB_d^{-} = 0.81 \bB^P_d + 0.58 \bB^T_d
\end{array}
\ee
\vspace{0.3cm}

We have assumed so far that $\langle\cL^2\bB_d^T\rangle$ and $\langle\cL^2\bB_d^P\rangle$ belong to the vector space generated by $\bB_d^T$ and $\bB_d^P$. In order to check the validity of this hypothesis, we show some comparisons in  figure~\ref{fig:err_dipole}, using profiles normalized by their maximum value: in (a) the axial profiles of the radial component of $\langle\cL^2\bB_d^T\rangle$ and $\langle\cL^2\bB_d^P\rangle$
are compared to the equivalent profile for $\bB_d^P$. One can see that the overlap is very good. In (b), the radial profiles of the azimuthal component of $\langle\cL^2\bB_d^T\rangle$ and $\langle\cL^2\bB_d^P\rangle$ are compared to the radial profile for $\bB_d^T$. In the case of the $\alpha$-effect ($\langle\cL^2\bB_d^P\rangle$ profile), the overlap is very good again. In the case of the expulsion effect ($\langle\cL^2\bB_d^T\rangle$ profile), there is a discrepancy. We have computed the error to be about 10\%. Thus the $\Mtt$ term in the matrix represents the expulsion mechanism for the applied toroidal field within an error of 10 \%, while the other elements of the matrix can be considered as correct within less than 1 \%. 

\subsubsection{Generation of a quadrupole field.}\label{sec:quadrupole}

We follow the same procedure as for the dipole case. We start from $\langle B_{10}\rangle$ presented in figure~\ref{fig:BrBSup}(g), which was shown (equation (\ref{equ:loop_radial})) to be an eigenvector of the operator $\cL^2$, with a quadrupolar geometry and a negative eigenvalue. This mode is orthogonal to the dipole mode used in the previous section. We split this field into its toroidal -- 
$\bB_q^T$ -- and poloidal -- $\bB_q^P$ -- components (figure~\ref{fig:L_quadrupole}(a) and (c)). Figures~\ref{fig:L_quadrupole}(b) and (d) show the axisymmetric part of the fields  $\cL^2\bB_q^T$ and $\cL^2\bB_q^P$. Again, the resulting fields are of opposite sign with respect to  $\bB_q^T$ and $\bB_q^P$ and have very similar topology. We define the matrix $M$ as in the dipole case, yielding:

\be 
M_q(T\mbox{\footnotesize{\it $\frac{1}{2}$}}) = -10^{-4} \bmat 10&12\\12&11 \emat 
\ee

This matrix has two real eigenvalues of opposite sign: $\lambda_q^+=2.10^{-4}$ and $\lambda_q^-=-23.10^{-4}=-1/430$. The $T\frac{1}{2}$ flow can thus also sustain an axial quadrupole dynamo, with a threshold $\Rm^c=1/\sqrt{\lambda_q^+}=80$. The two corresponding eigenmodes have the following structure:

\be
\begin{array}{c}
\bB_q^{+} = 0.72 \bB^P_q - 0.69 \bB^T_q\\
\\
\bB_q^{-} = 0.69 \bB^P_q + 0.72 \bB^T_q
\end{array}
\ee

As in the dipole case, the error made assuming that $\langle\bB_2\rangle$ belongs to the vector space generated by $\bB^P_q$ and $\bB^T_q$ can be estimated by comparing the profiles of $\langle\cL^2\bB_2^P\rangle$ and $\langle\cL^2\bB_2^T\rangle$ to $\bB^P_q$ and $\bB^T_q$. We find here also that the only component that does not overlap correctly is the one corresponding to the expulsion of the azimuthal applied field -- $\Mtt$ -- with an error of about 10\%.

\subsubsection{Transverse dipole.}\label{sec:transverse}

Until now, we have only considered axisymmetric fields. However, since the $T\frac{1}{2}$ flow presents several analogies with the Roberts flow, it would be interesting to study the possibility of generating a transverse dipole (perpendicular to the columns), as observed in the Karlsruhe dynamo~\cite{Karlsruhe}. We follow the same strategy as before: a uniform transverse field $\bB_0=B_0\bex$ is applied to the flow and we compute the fields $\bB_k$ obtained after $k$ iterations. They rapidly converge towards a stable structure. After 8 iterations, the fields $\bB_k$ and $\cL^2\bB_k$ have an overlap close to 100\%, and the eigenvalue is negative, $\gamma=-1/400$. Figure \ref{fig:transverse_mode} shows two cuts of the field $\bB_8$. 

In order to use the same matrix analysis as in the axisymmetric cases, one must find a way to decompose $\bB_8$ into a couple of vectors generating a subspace closed under $\cL^2$. No poloidal/toroidal decomposition can be made here since the mode is not axisymmetric, but  symmetry considerations can be used, based on the following observation: in section \ref{sec:dipole} and \ref{sec:quadrupole} the toroidal/poloidal decomposition of the eigenvector also resulted in separating the symmetric and antisymmetric parts of the eigenvector under the reflection symmetry with respect to the plane $z=0$. In what follows, we will call \Sz~this symmetry.  One can then notice in figure~\ref{fig:transverse_mode} that the $x$ component of the field is symmetric under \Sz, while the $y$ component is antisymmetric. The system being finite and the exterior being insulating, the $x$ and $y$ components are each associated to an axial component having the same symmetry: $B_x$, symmetric with respect to \Sz, is associated with a symmetric axial component, $B_z^s$; $B_y$, antisymmetric, is associated with the antisymmetric part of the axial component, $B_z^a$. We thus define the vectors  $\bB^0_x=B_x\bex+B^s_z\bez$ and $\bB^0_y=B_y \bey+B^a_z\bez$ as the generating vectors.

Applying $\cL^2$ to these vectors, one observes that, in this case too, the resulting vectors can be expressed as linear combinations of $\bB^0_x$ and $\bB^0_y$. $\cL^2$ can therefore be expressed in this basis as the matrix :

 \be 
 M_t(T\mbox{\footnotesize{\it $\frac{1}{2}$}}) = -10^{-4} \bmat 15&13\\20&14 \emat 
 \ee
 
 This matrix has a positive eigenvalue $\lambda_t^+=1.6~10^{-4}$, showing that the $T\frac{1}{2}$ flow is also able to sustain a transverse dipole, with a threshold equal to 79, very close to the one of the axial quadrupole. 
 
 In a $T\frac{1}{2}$ flow, we thus observed that several dynamo modes can be sustained when $\Rm\simeq 80$. This result is consistent with other numerical studies in thermal convection~\cite{aubert}, which have shown that various dynamo solutions can coexist in the same region of the parameter space. This kind of behavior has also been observed in the VKS dynamo experiment~\cite{inversions}.

\subsection{Dynamo mechanisms in the $T1$ flow.}\label{dynamo_T1}

\subsubsection{Symmetry considerations.}

The analytical expression of the $T1$ flow is given in equation (\ref{equ:T1flow}). Compared to the $T\frac{1}{2}$ flow (equation (\ref{equ:Tdemiflow})), the only difference is that the azimuthal and axial components are now respectively symmetric and antisymmetric under ${\cal S}_z$. $T1$ is a superposition of two $T\frac{1}{2}$ flows, one in each half-cylinder, symmetric with respect to ${\cal S}_z$. They have opposite helicity, since their axial component is reversed, while the rotation of the columns is unchanged. The strong similarity between the $T\frac{1}{2}$ and $T1$ flows indicates that the same mechanisms, $\alpha$ effect and expulsion, will take place. Figure \ref{fig:modes_T1} compares the schematic induction mechanisms for the $\alpha$ effect in both types of flow. In (a), the case of the $T\frac{1}{2}$ flow is recalled, where the induced current ($\mathbf{j}=\sigma\alpha\bB$) is parallel to the applied field with the same sign, since helicity $\cal{H}$ is negative in this flow and the $\alpha$ coefficient is proportional to $-\cal{H}$. In (b), it is shown that in the $T1$ flow, applying a symmetric toroidal field results in a symmetric poloidal field, corresponding to a quadrupolar geometry. In (c), we show that the dipolar geometry (antisymmetric poloidal field) is obtained by applying an antisymmetric toroidal field. 

\subsubsection{Search of axisymmetric dynamo modes.}\label{T1_axi}

{\bf Dipolar mode:}\\

The symmetry considerations of the previous paragraph suggest that a dipolar mode can be obtained in the $T1$ flow by an iterative application of $\cL$ to an initial toroidal field antisymmetric with respect to ${\cal S}_z$. We chose an initial field of the form $\bB_0=B_0 \sin(\pi\frac{z}{H})\bet$. The iterations converge rapidly towards a mode whose toroidal and poloidal parts, once the azimuthal average is done, form a basis which is closed under the action of $\cL^2$. The corresponding matrix is:

\be 
M_d(T1)= -10^{-4} \bmat 13&16\\8&12 \emat 
\ee 

It has two negative eigenvalues $\lambda^d_1=-24$ $10^{-4}$ and $\lambda^d_2=-1.2$ $10^{-4}$, so that with the current parameters ($d/r=0.4$, $\xi=1.25$ and $n=4$), the $T1$ flow is unable to sustain an axial dipole dynamo mode. \\

{\bf Quadrupolar mode:}\\

In this case, the symmetry study shows that the initial azimuthal field must be symmetric under ${\cal S}_z$. Two choices can be tested: an azimuthal $z$-independent field or a radial field of the form  $\bB_0=B_0
 \frac{a}{r} \cos(\pi z/H) \ber$. In both cases the iterations converge towards fields with identical structures. As previously, the toroidal and poloidal components of the field obtained after 10 iterations define the basis for the quadrupole vector space. In this basis, the expression of $\cL^2$ is:
 
 \be 
 M_q(T1)= -10^{-4} \bmat 11&11\\9.1&9.4 \emat 
 \ee
 
 Again this matrix has two negative eigenvalues $\lambda_1^q=-20.10^{-4}$ and $\lambda_2^q=-0.2.10^{-4}$. \\
 
It is interesting to note that in all the cases where a dynamo was possible, the corresponding matrix ($M_d(T\mbox{\footnotesize{\it $\frac{1}{2}$}})$, $M_q(T\mbox{\footnotesize{\it $\frac{1}{2}$}})$ and $M_t(T\mbox{\footnotesize{\it $\frac{1}{2}$}})$) presented the characteristics that the product of the non-diagonal terms was larger than the product of the diagonal terms. One can easily show that this is a necessary and sufficient condition for a 2$\times$2 matrix with negative diagonal coefficients in order to have a positive eigenvalue. And this condition physically corresponds to the facts that the expulsion mechanism (diagonal terms) is weaker than the $\alpha$ mechanism (non-diagonal terms). On the contrary, in the case of ($M_d(T1)$) and ($M_q(T1)$), this condition is not met. In the quadrupolar case, both products are very close, resulting in a small value for  $\vert\lambda_2^q\vert$ (close to $10^{-5}$). We can therefore expect that by increasing the $\xi$ parameter (reducing the rotational component of the flow, responsible for the expulsion effect), a quadrupolar dynamo could be observed in the $T1$ flow.\\
 
{\bf Transverse mode:}\\
 
In the case of the \Tdemi , we have iterated the $\cL$ operator starting from a uniform transverse field and then defined the basis vectors using as a criterion the behavior of the cartesian components of the converged mode under the \Sz~symmetry. We tried to follow the same procedure in the case of the $T1$ flow, but it turned out that the converged mode is formed of two dipolar structures at 90 degrees from each other, having the same behavior with respect to the \Sz~symmetry. Therefore, it is not possible to use this symmetry to construct the basis vectors. And indeed, we were not able to find any basis that would be closed under the action of $\cL^2$. This shows a limit of our method.

\subsection{Generalization for other values of the poloidal/toroidal ratio.}

\subsubsection{General expression for the matrix $M$.}

Until now, all our studies have been based on the same flow parameters ($d/r=0.4$, $\xi=1.25$ and $n=4$). As we noticed in section~\ref{T1_axi} that a change in the axial/rotational ratio ($\xi$ parameter) might allow the $T1$ flow to sustain a quadrupolar dynamo, we redefine the $\cL$ operator as a linear function of the poloidal and toroidal components of the flow. We normalize the components $\bv^A$ and $\bv^R$ of the velocity field: $\maxx(\bv^A)=1$ and $\maxx(\bv^R)=1$ and rewrite equation~(\ref{equ:velocity}) as $\disp \bV=V_A \bv^A + V_R \bv^R$ where $V_A$ and $V_R$ represent the maximum amplitude of each component. Let $\cL_A$ be the induction operator when $\bV=\bv^A$ and $\cL_R$ be the induction operator when $\bV=\bv^R$; $\cL$ can then be written as the linear combination:

\be 
\disp  \cL=V_A \cL_A + V_R \cL_R\;\;\;,
\ee

yielding:

\be 
\disp \cL^2=V^2_A\cL_A\cL_A + V_AV_R (\cL_R\cL_A+\cL_A\cL_R) + V^2_R \cL_R \cL_R\;\;\;,
\label{decompL}
\ee

so that the 4 matrix elements $M_{ij}$ can also be written as quadratic forms of $V_A$ and $V_R$:

\be  \label{matrix_element}
\disp M_{ij}=a_{ij}V^2_A + b_{ij}V_RV_A+ c_{ij}V^2_R.
\ee

As we studied the induction mechanisms, we noticed that the $\alpha$ effect and the expulsion behave differently under a reversal of the axial pumping ($V_A \rightarrow -V_A$) or of the columns rotation  ($V_R \rightarrow -V_R$). More precisely, it was observed that the expulsion effect is independent of these sign changes, whereas the $\alpha$ effect transforms as the product $V_AV_R$. These observations allow to eliminate some terms in equation (\ref{matrix_element}), yielding:
 \be 
 \disp  M(V_R,V_A) = \bmat  \disp a V_R^2 + bV_A^2& \disp c V_RV_A\\ \disp dV_RV_A& \disp e V_R^2 + f V_A^2\emat 
\label{matvpvt}
\ee

\subsubsection{Positive eigenvalue diagram for the \Tdemi .}

In the case of the \Tdemi, the value of these 6 coefficients are computed using successively the dipole basis, the quadrupole basis and the transverse basis obtained respectively in sections~\ref{sec:dipole}, ~\ref{sec:quadrupole} and ~\ref{sec:transverse}. The corresponding expressions for the matrix $M$ are the following:\\

{\bf Axial dipole:}
\be 
M_{\text{d}}(V_A,V_R) =
-10^{-4} \bmat 22 V_A^2 + 6 V_R^2&46 V_AV_R\\18V_AV_R&14(V_A^2 +V_R^2) \emat
\ee
{\bf Axial quadrupole:}
\be
M_{\text{q}}(V_A,V_R) =
 -10^{-4} \bmat 23 V_A^2 + 5 V_R^2&25 V_AV_R\\25V_AV_R&11(V_A^2 + V_R^2) \emat 
 \ee
{\bf Transverse dipole:}
\be M_{\text{t}}(V_A,V_R) = 
-10^{-4} \bmat 22 V_A^2 + 13V_R^2&29 V_AV_R\\46V_AV_R&16V_A^2 + 14V_R^2 \emat 
\ee

From these expressions, the largest eigenvalue $\lambda_{max}$ of each matrix can be computed as a function of $V_A$ and $V_R$. The result is shown in figure~\ref{fig:Tdemi_diag}. In order to facilitate the reading of these plots, when $\lambda_{max}$ was negative (no dynamo), we artificially set its value to zero. For each case, two regions are evidenced: the first one, for which $\lambda_{max}>0$, corresponds to the possibility to observe a dynamo for the considered geometric parameters ( $d/r=0.4$ and $n=4$) with a threshold $\Rm^c=1/\sqrt{\lambda_{\maxx}}$. The second, for which $\lambda_{max}<0$, corresponds to the case where no dynamo instability can take place. Figures (a), (b) and (c) show that the three dynamo modes coexist in the same region, in the neighborhood of the line $V_A=V_R$. This is not surprising, since when one of the velocity component dominates the other, the expulsion mechanism is more important than the $\alpha$ effect, which needs both components together.

For a given couple $(V_A,V_R)$, one can observe that $\lambda_q$ is always larger than $\lambda_d$ (corresponding to a lower threshold). In the same way, the transverse dipole always has a lower threshold than the axial dipole. A summary of the predominance of the different modes is given in figure~\ref{fig:Tdemi_diag}(d), where one can see that the transverse dipole is favored by higher ratio axial/rotational, whereas the quadrupole is favored by a lower ratio.

\subsubsection{Positive eigenvalue diagram for the $T1$ flow.}

We performed the same analysis for the $T1$ flow. The matrix $M$ has the following expressions:

{\bf Axial dipole:}
 \be 
 M_{\text{d}}(V_A,V_R) =
-10^{-4} \bmat 16 V_A^2 + 8 V_R^2&33 V_AV_R\\16V_AV_R&12(V_A^2 +V_R^2) \emat
\ee
{\bf Axial quadrupole:}
 \be
M_{\text{q}}(V_A,V_R) = 
-10^{-4} \bmat 20 V_A^2 + 5 V_R^2&25 V_AV_R\\20V_AV_R&11V_A^2 + 8V_R^2 \emat 
\ee

Figure~\ref{fig:T1_diag} shows the evolution of $\lambda_{max}$ in the case of the quadrupole mode. It confirms the assumption made in section~\ref{T1_axi}, namely that starting from the value $\xi=1.25$ studied in that section (cross in figure~\ref{fig:T1_diag}), one could evolve the system towards a dynamo state by reducing the intensity of the rotation (increasing $\xi$, corresponding in figure~\ref{fig:T1_diag} to a displacement towards the top left corner).

On the other hand, there is no value of the couple $(V_A,V_R)$ for which the axial dipole can be sustained by the $T1$ flow. Some studies have shown~\cite{cardin1} that in order for a $T1$-type flow to sustain an axial dipole, the presence of differential rotation, absent in our study, would be necessary.

 \section{Concluding remarks.}
   
A better understanding of the MHD induction mechanisms in a given system can help to build dynamo cycles. In the first part of this study, we have identified two mechanisms, related respectively to the $\alpha$-effect and to expulsion by vortices. We expressed these mechanisms in terms of the induction operator $\cL^2$ and, using a poloidal/toroidal decomposition of the eigenvectors of $\cL^2$, we were able to perform a matrix analysis leading to the determination of self-sustained magnetic modes. The benefit is to give a complete description of the modes whose interaction lead to dynamo action. In regards to natural or experimental conditions, it helps understand which features of the velocity field favor or hinder dynamo action. In addition, as already noticed in other studies~\cite{vkgalpha}, the competition between the $\alpha$-effect, favorable to the dynamo, and the expulsion effect, that works against it, can be monitored by the ratio of poloidal to toroidal components of the velocity field (in our case the axial to rotational ratio $V_A/V_R$). Our studies show that a positive feed-back requires a comparable amplitude for rotation and pumping. In the case of the \Tdemi~(for 8 columns and an aspect ratio of 0.4), we observed that both axisymmetric modes (dipolar and quadrupolar) can be sustained, with a threshold of the order of $\Rm=100$. We also showed that a transverse dipole mode can exist, as one could expect, because of the analogy between our flow and the Karlsruhe dynamo. On the other hand, the $T1$ flow  can only sustain a quadrupolar mode, when the pumping amplitude is larger than the rotation amplitude. For the axial dipole mode, the poloidal to toroidal conversion seems too weak to compensate the strong expulsion of the azimuthal field.\\
  
Coming back to the Earth's case, our model system leads to several observations which may be relevant. An $\alpha^2$ dynamo process relies on the helicity contained in Busse's columns~\cite{Busse}, but Eckman pumping 
would give a very weak source of axial motion, since the Eckman number $E$ is of the order of $10^{-15}$ and the ratio of the axial flow to the rotational flow scales like $E^{1/2}$~\cite{nataf}. 
Another source of axial velocity could be by the $\beta$-effect due to the curvature of the core-mantle boundary -- note that in this case the axial flow is in phase with the radial flow rather than with the vorticity~\cite{cardin1}.  A large scale dipole field could also be generated from an $\alpha-\omega$ dynamo. It would require differential rotation as provided, for instance, by zonal winds~\cite{aubert2, gilet} or super-rotation effects as observed in a recent laboratory experiment~\cite{superrotation} These ingredients could in principle be added to the model studied here, and the procedure used to determine which dynamo modes (dipole, quadrupole or other) are likely to exist for a given range of Reynolds numbers. As proposed for geomagnetism~\cite{MacFadden}, and recently observed in the VKS experiment~\cite{inversions}, the close proximity of dynamo modes may be essential for the development of dynamical regimes. \\

\noindent{\bf Acknowledgements}\\
We acknowledge useful discussions with M. Bourgoin, R. Avalos-Zu\~niga, P. Cardin, F. Plunian, and N. Schaeffer. This work was partially supported by the \'Emergence Program of the Rh\^one-Alpes Region (France). 




\newpage
\rightline{\bf Volk {\it et al.}, Figure 1}
\vspace*{10mm}
\begin{figure}[h]
\centerline{\includegraphics[width=12cm]{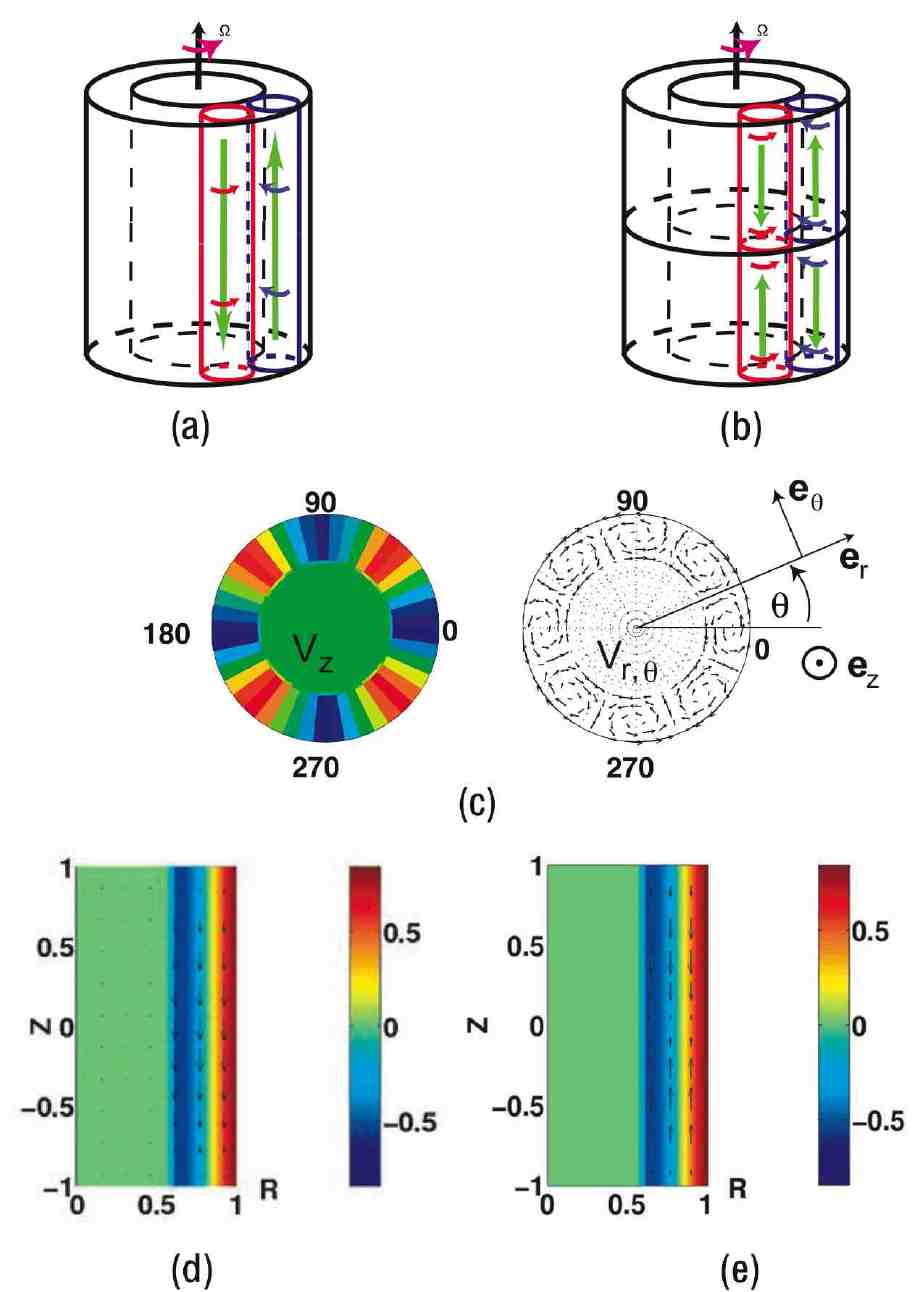}}
\caption{ Geometry investigated. (a) Sketch of the column arrangement for the $T\frac{1}{2}$ flow. (b) Column arrangement for the $T1$ flow. (b) Cut of the $T\frac{1}{2}$ flow in the plane $z=0$ (colors correspond to the vertical flow, arrows correspond to the flow in the plane $z=0$) (d) Cut of the $T\frac{1}{2}$ flow in the plane $\theta=0$. (e) Cut of the $T1$ flow in the plane $\theta=0$.}
\label{fig:geometry}
\end{figure}

\newpage
\rightline{\bf Volk {\it et al.}, Figure 2}
\vspace*{10mm}
\begin{figure}[h]
\centerline{\includegraphics[width=13cm]{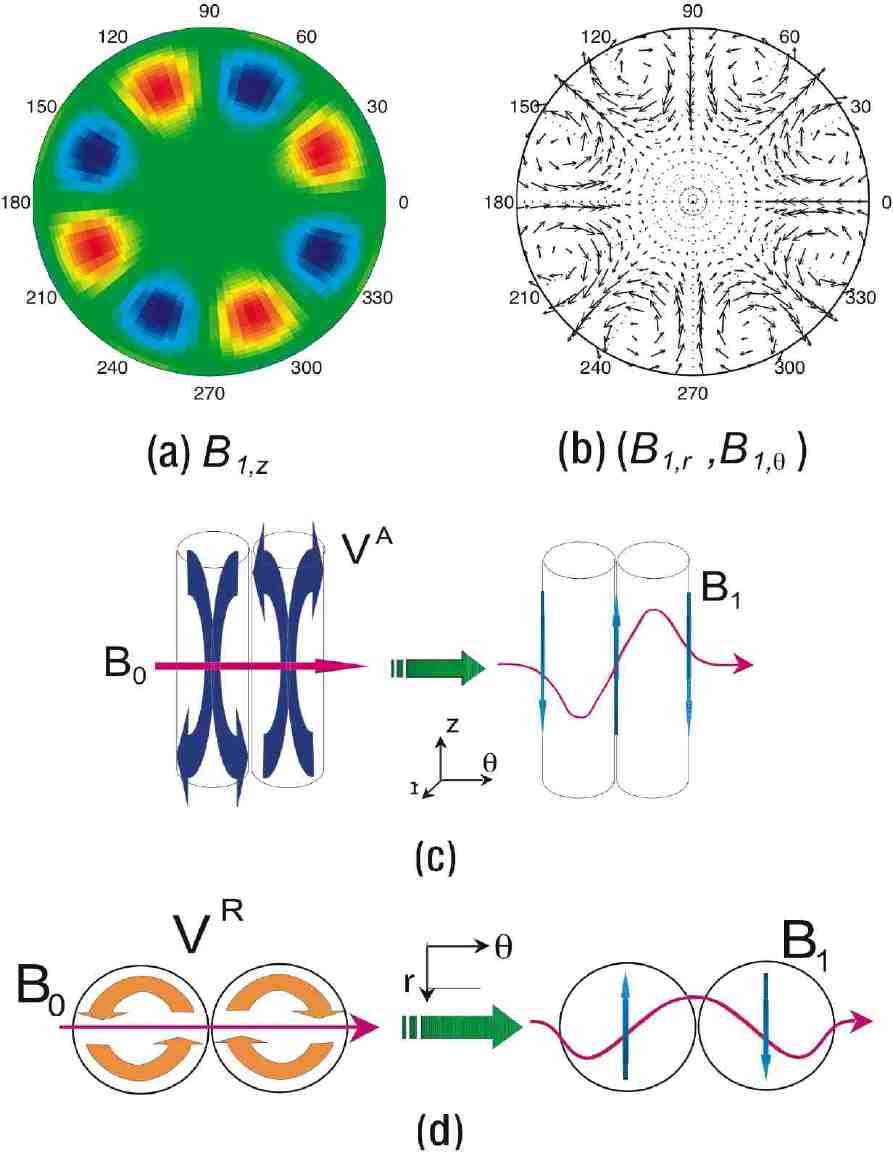}}
\caption{$T\frac{1}{2}$: magnetic field induced at first order for an applied toroidal field $B_{0,\theta}$. Fields cuts are shown in the $z=0$ mid-plane. (a) induced field along the axial direction, $B_z$; (b) transverse components $(B_r, B_\theta)$. Associated distortion of the applied field lines: (c) axial induced field from axial stretching between columns. (d) radial induced field from the rotational flow.}
\label{fig:BthetaB1}
\end{figure}

\newpage
\rightline{\bf Volk {\it et al.}, Figure 3}
\vspace*{10mm}
\begin{figure}[h]
\centerline{\includegraphics[height=16cm]{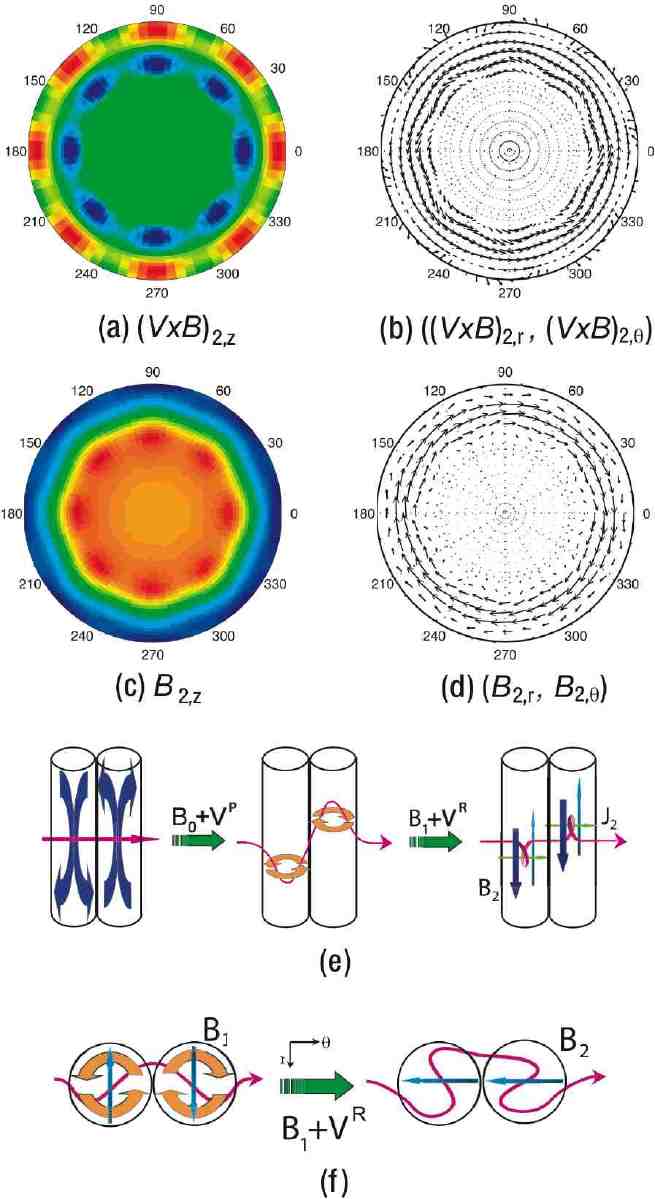}}
\caption{$T\frac{1}{2}$: magnetic field induced at second order for an applied toroidal field $B_{0,\theta}$. (a): vertical component of the electromotive force $e_{2,z}$, shown in the mid plane $z=0$, (b): transverse components $(e_{2,r}, e_{2,\theta})$ in the same $z=0$ plane; (c) corresponding induced magnetic field $B_{2,z}$ and (d): $(B_{2,r}, B_{2,\theta})$; (e): schematics of the cooperative induction effect from a pair of neighboring columns, generation of $B_{2,z}$; (f): associated expulsion mechanism, generation of $B_{2,\theta}$. }
\label{fig:BthetaB2}
\end{figure}

\clearpage
\newpage
\rightline{\bf Volk {\it et al.}, Figure 4}
\vspace*{10mm}
\begin{figure}[h]
\centerline{\includegraphics[width=15cm]{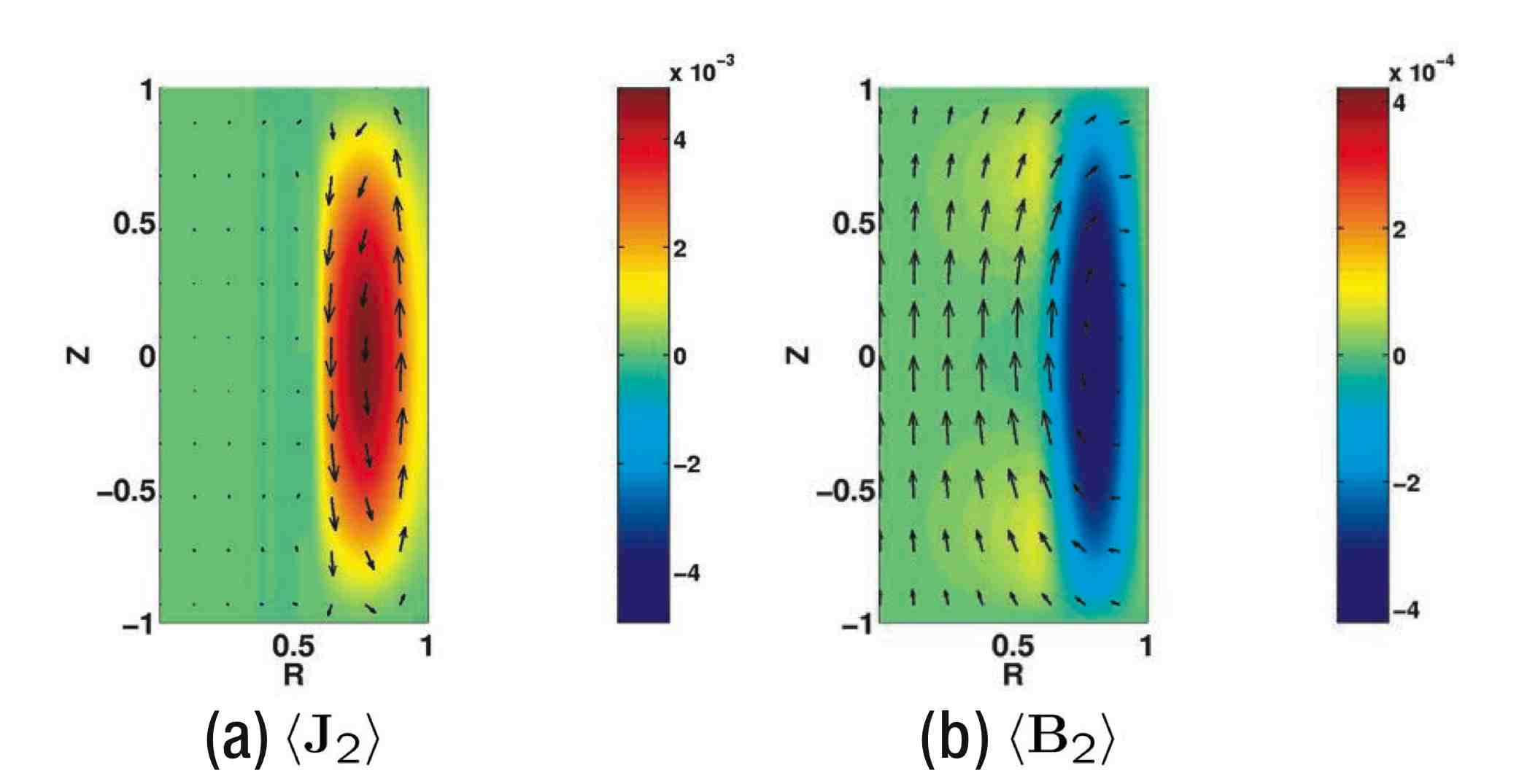}}
\caption{$T\frac{1}{2}$: azimuthal average of the fields induced at second order for an applied toroidal field $B_{0,\theta}$. Azimuthal averages of the fields are shown in an $(r,z)$ plane, with poloidal components indicated by arrows and azimuthal component in color scale. (a): current $\langle {\mathbf J}_2 \rangle$; (b): corresponding magnetic field $\langle \bB_2 \rangle$. }
\label{fig:BthetaB2moy}
\end{figure}

\clearpage
\newpage
\rightline{\bf Volk {\it et al.}, Figure 5}
\vspace*{10mm}
\begin{figure}[h]
\centerline{\includegraphics[width=15cm]{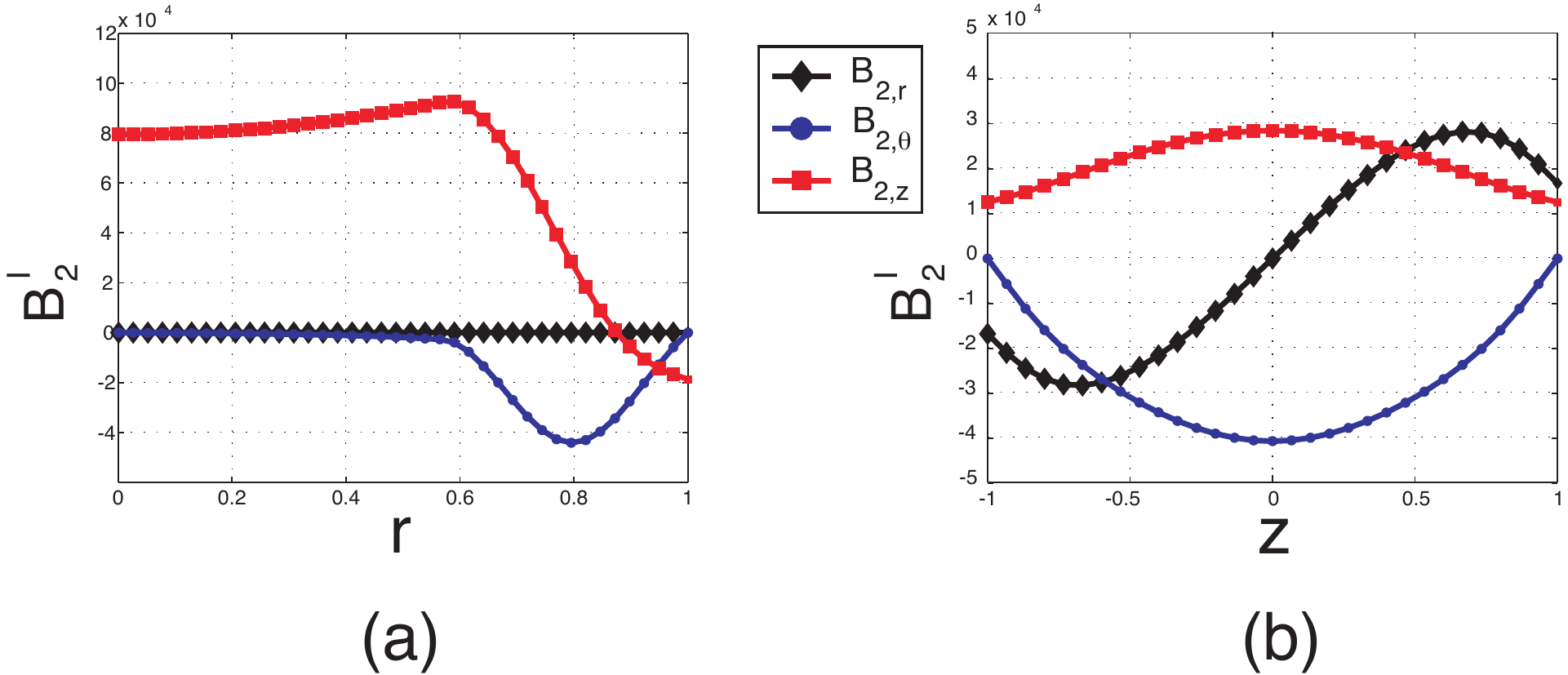}}
\caption{$T\frac{1}{2}$: spatial profiles of $\langle \bB_2 \rangle$. (a): variation with $r$ in the $z=0$ plane; (b): axial variation at $r=0.8$ (the center of the columns). The magnitude of the applied field is 0.25 in the center of the columns.}
\label{fig:BthetaB2prof}
\end{figure}

\newpage
\rightline{\bf Volk {\it et al.}, Figure 6}
\vspace*{10mm}
\begin{figure}[h]
\centerline{\includegraphics[height=15cm]{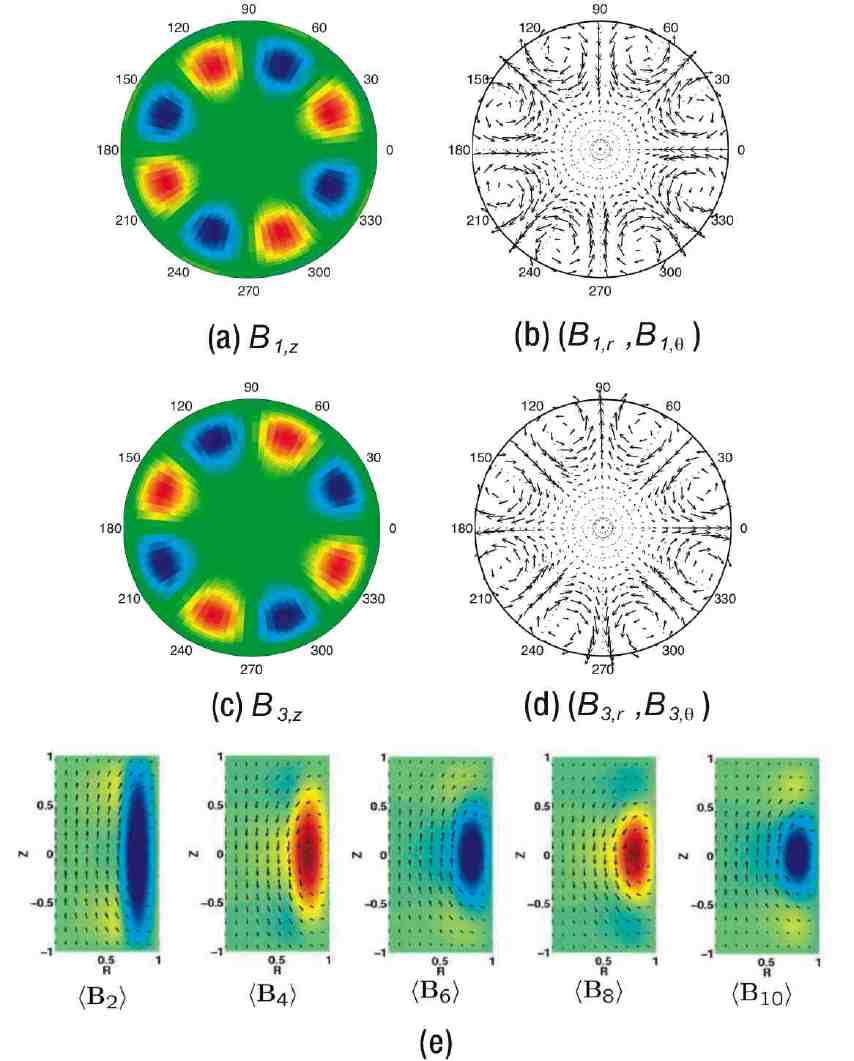}}
\caption{$T\frac{1}{2}$: magnetic field induced at higher orders for an applied toroidal field $B_{0,\theta}$. (a,b): axial and transverse components of  $\bB_1$; (c,d): corresponding plots for $\bB_3$; (e): evolution of the azimuthal average $(\langle \bB_2 \rangle, \langle \bB_{2k} \rangle, \ldots \langle \bB_{10} \rangle)$}
\label{fig:BthetaBSup}
\end{figure}

\newpage
\rightline{\bf Volk {\it et al.}, Figure 7}
\vspace*{10mm}
\begin{figure}[h]
\centerline{\includegraphics[width=15cm]{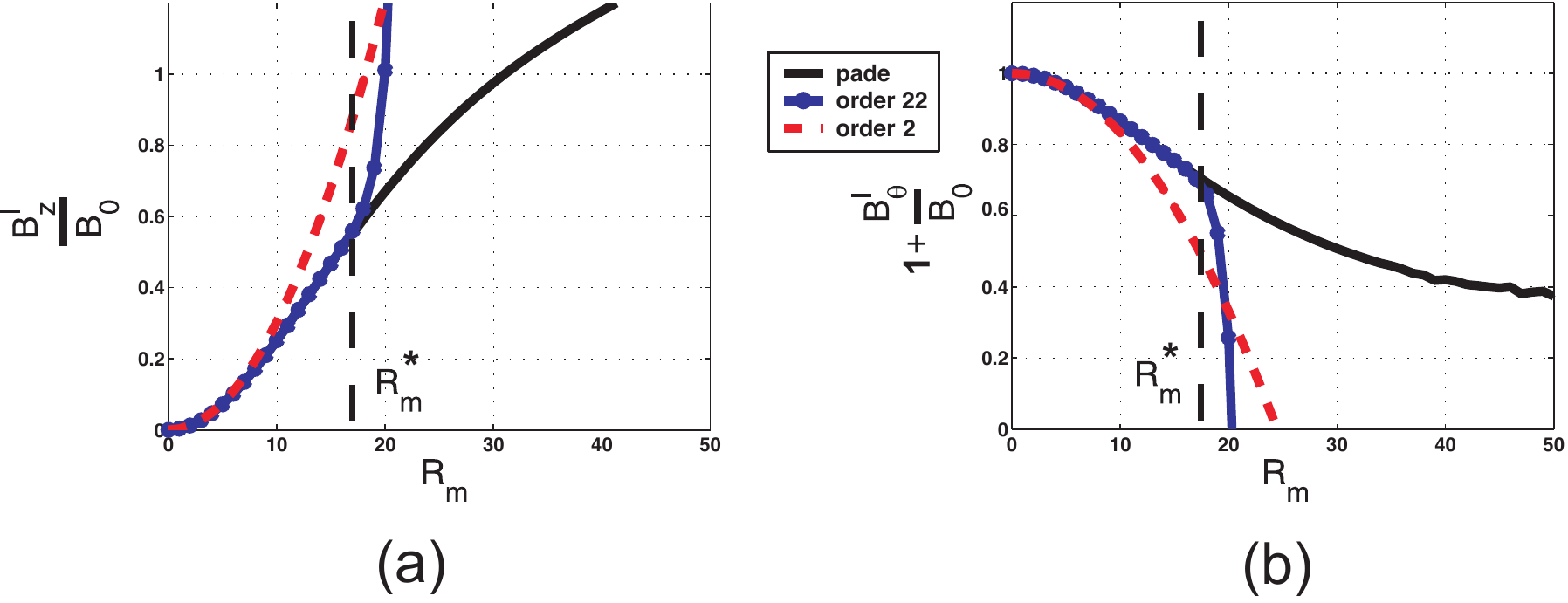}}
\caption{$T\frac{1}{2}$: evolution with $\Rm$. (a): axial field in the center of the annulus, at $(r=0, z=0)$; (b): toroidal field in the columns, at  $(r=0.8, z=0)$. $\Rm^\ast$ is the radius of convergence of the integer series.
The dashed line corresponds to the summation stopped at order 2, the continuous line with dots to the summation up to order 22 and the continuous line to the summation using Pad\'e approximants. }
\label{fig:BthetaBSupProfiles}
\end{figure}

\newpage
\rightline{\bf Volk {\it et al.}, Figure 8}
\vspace*{10mm}
\begin{figure}[h]
\centerline{\includegraphics[width=15cm]{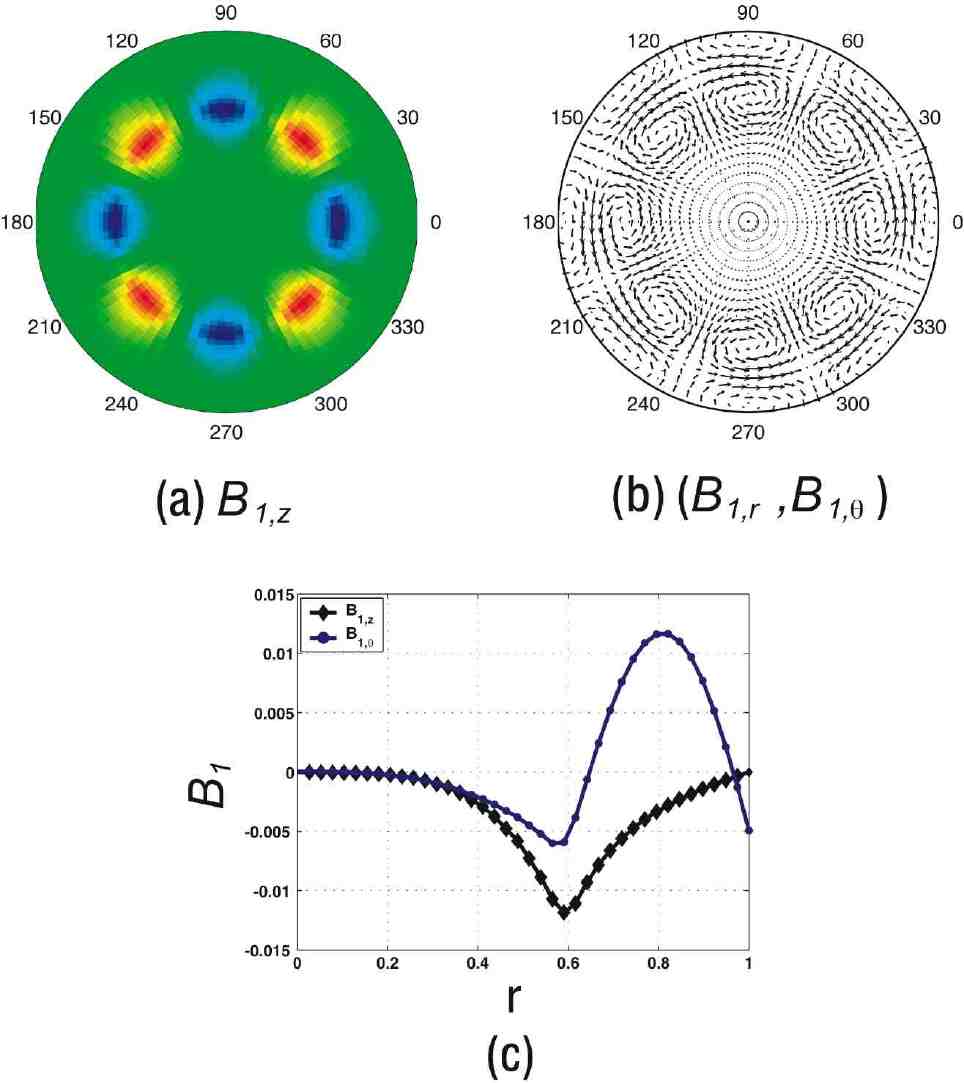}}
\caption{$T\frac{1}{2}$: magnetic field induced at first order for an applied radial field $B_{0,r}$. Field cuts are shown in the $z=0$ mid-plane. (a) induced field along the axial direction, $B_z$; (b) transverse components $(B_r, B_\theta)$. (c) Radial profiles of the induced fields, at $(z=0, \theta=0)$.}
\label{fig:BrB1}
\end{figure}


\newpage
\rightline{\bf Volk {\it et al.}, Figure 9}
\vspace*{10mm}
\begin{figure}[h]
\centerline{\includegraphics[width=11cm]{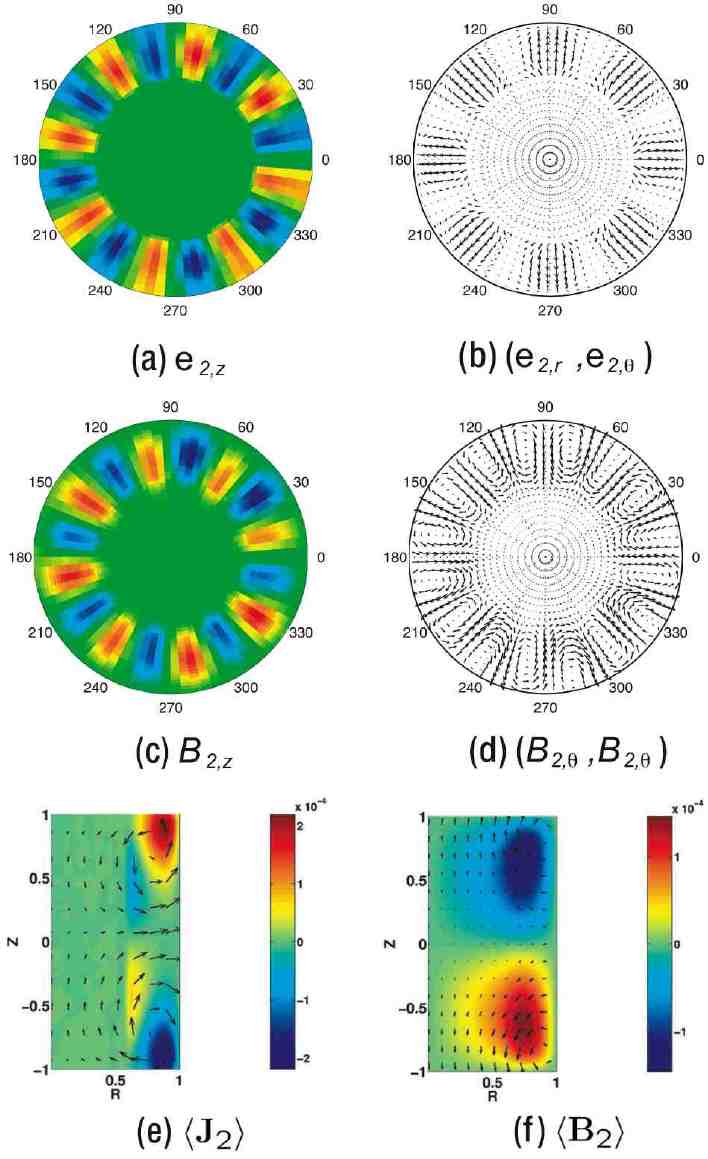}}
\caption{$T\frac{1}{2}$: induction at second order for an applied radial field $B_{0,r}$. (a,b) electromotive force; (c,d) induced field $\bB_2$; (e,f) meridian view, for the azimuthally averaged fields $\langle \bJ_2 \rangle$ and $\langle \bB_2 \rangle$. }
\label{fig:BrB2}
\end{figure}

\newpage
\rightline{\bf Volk {\it et al.}, Figure 10}
\vspace*{10mm}
\begin{figure}[h]
\centerline{\includegraphics[width=10cm]{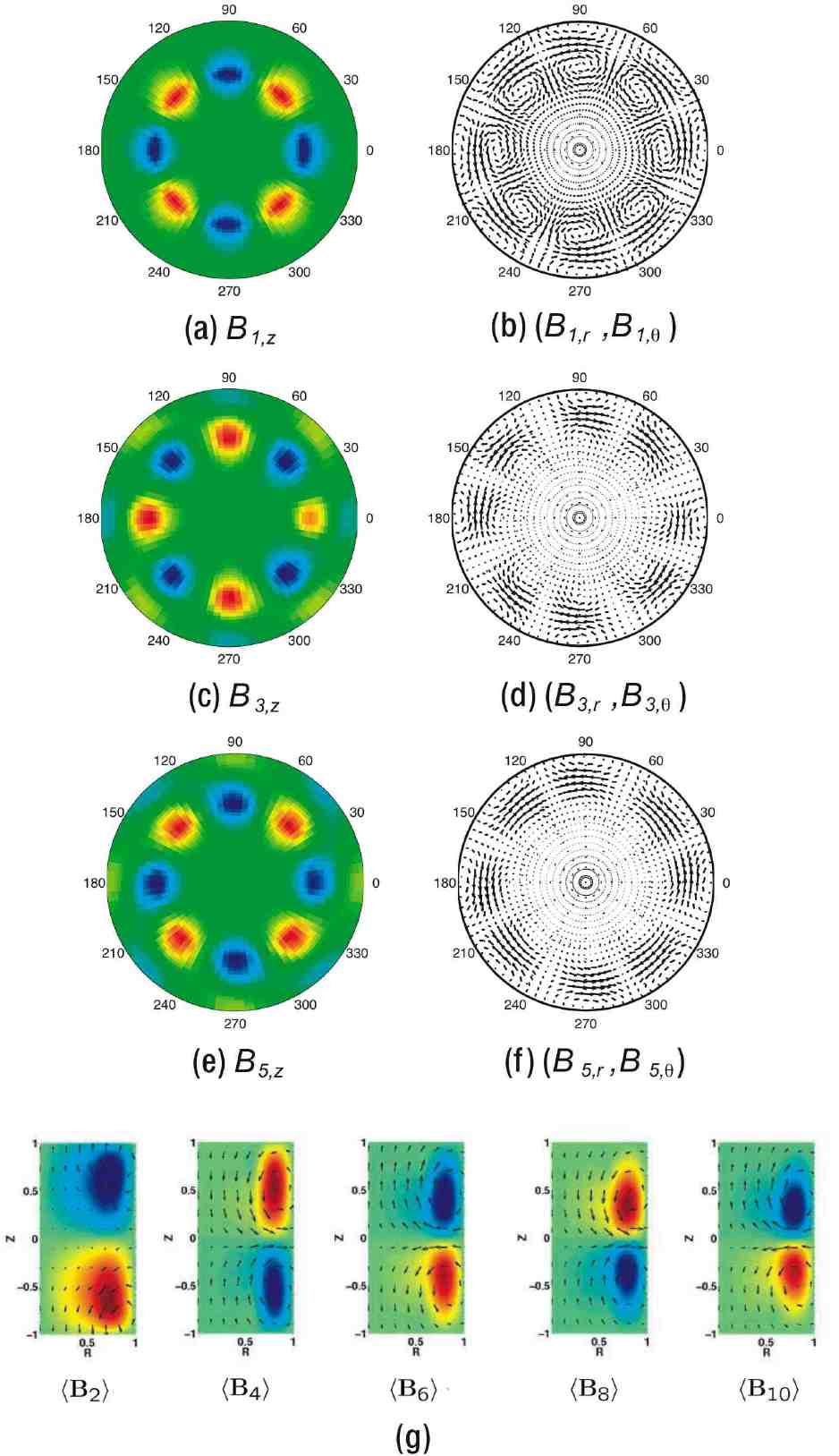}}
\caption{$T\frac{1}{2}$: induction at higher orders for an applied radial field $B_{0,r}$. (a-f): structure of the field induced at orders 1, 3 and 5, shown in the mid plane $z=0$. (g): structure of the azimuthal averaged $\langle \bJ_n \rangle$ of the fields induced at orders $n=2, 4, 6, 8, 10$.}
\label{fig:BrBSup}
\end{figure}

\newpage
\rightline{\bf Volk {\it et al.}, Figure 11}
\vspace*{10mm}
\begin{figure}[h]
\centerline{\includegraphics[width=16cm]{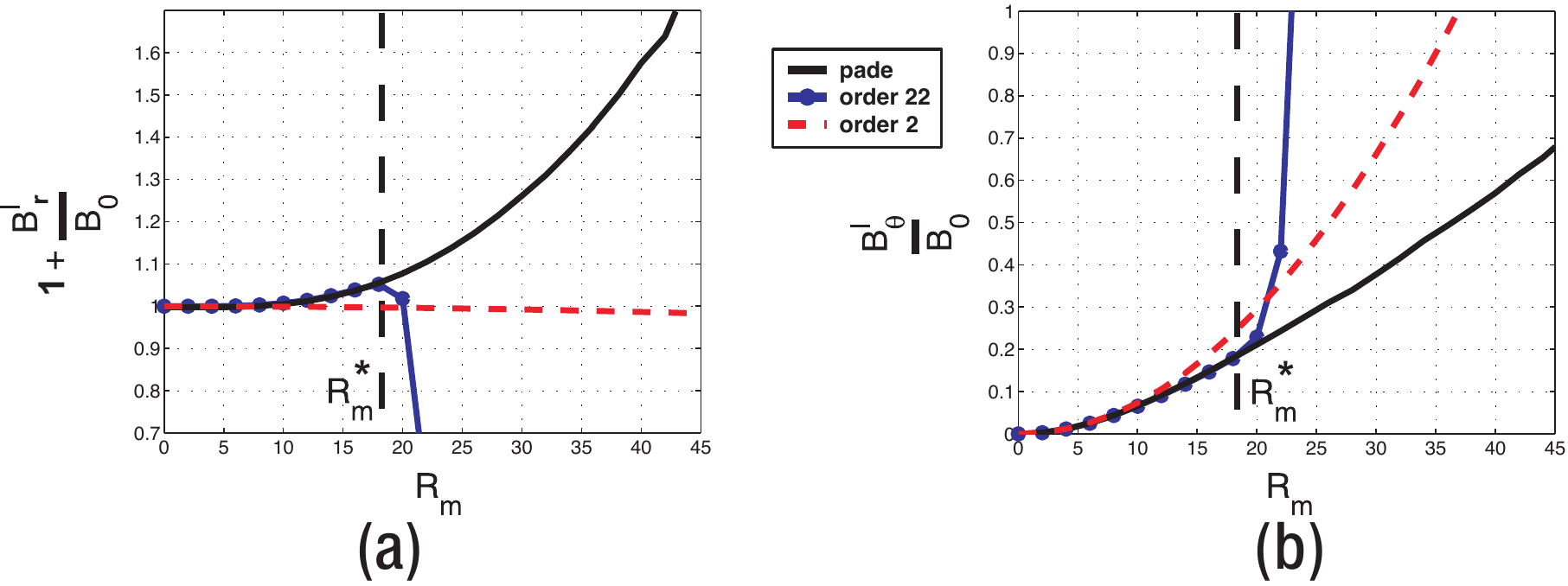}}
\caption{$T\frac{1}{2}$: induction at higher orders for an applied radial field $B_{0,r}$. Variation with $\Rm$ of the induced fields.  (a) radial induced component, sampled at $(r=0.7, z=0)$. (b) azimuthal induced component, sampled at $(r=0.7, z=-0.6)$. The dashed line corresponds to the summation stopped at order 2, the continuous line with dots to the summation up to order 22 and the continuous line to the summation using Pad\'e approximants.}
\label{fig:BrBSupProf}
\end{figure}

\newpage
\rightline{\bf Volk {\it et al.}, Figure 12}
\vspace*{10mm}
\begin{figure}[h]
\centerline{\includegraphics[width=15cm]{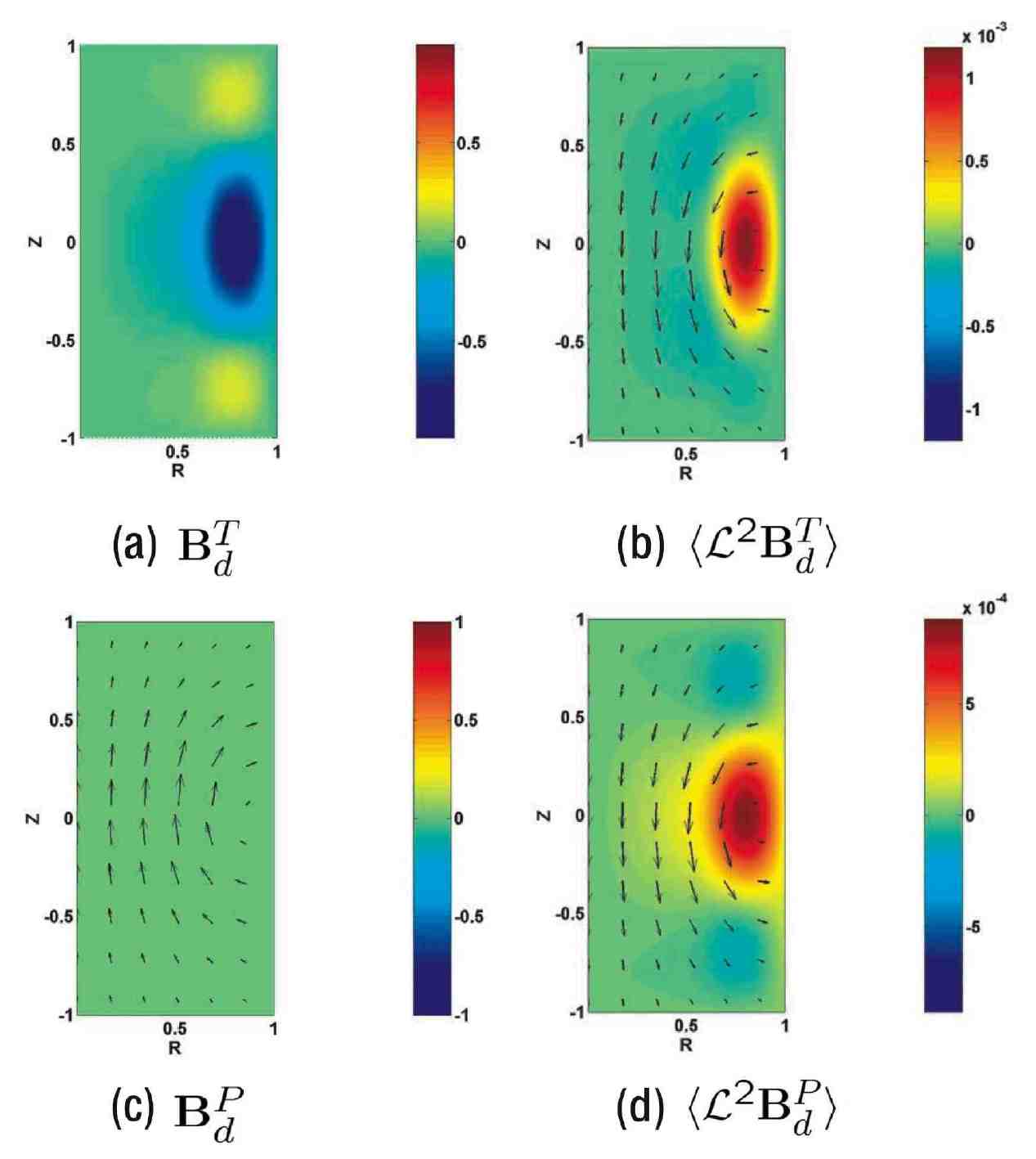}}
\caption{Dipolar mode: (a),(c) resp. toroidal and poloidal component of the dipole eigenvector. (b), (d), axisymmetric part of the fields obtained by applying $\cL^2$ to the vector fields (a) and (c).}
\label{fig:L_dipole}
\end{figure}

\newpage
\rightline{\bf Volk {\it et al.}, Figure 13}
\vspace*{10mm}
\begin{figure}[h]
\centerline{\includegraphics[width=15cm]{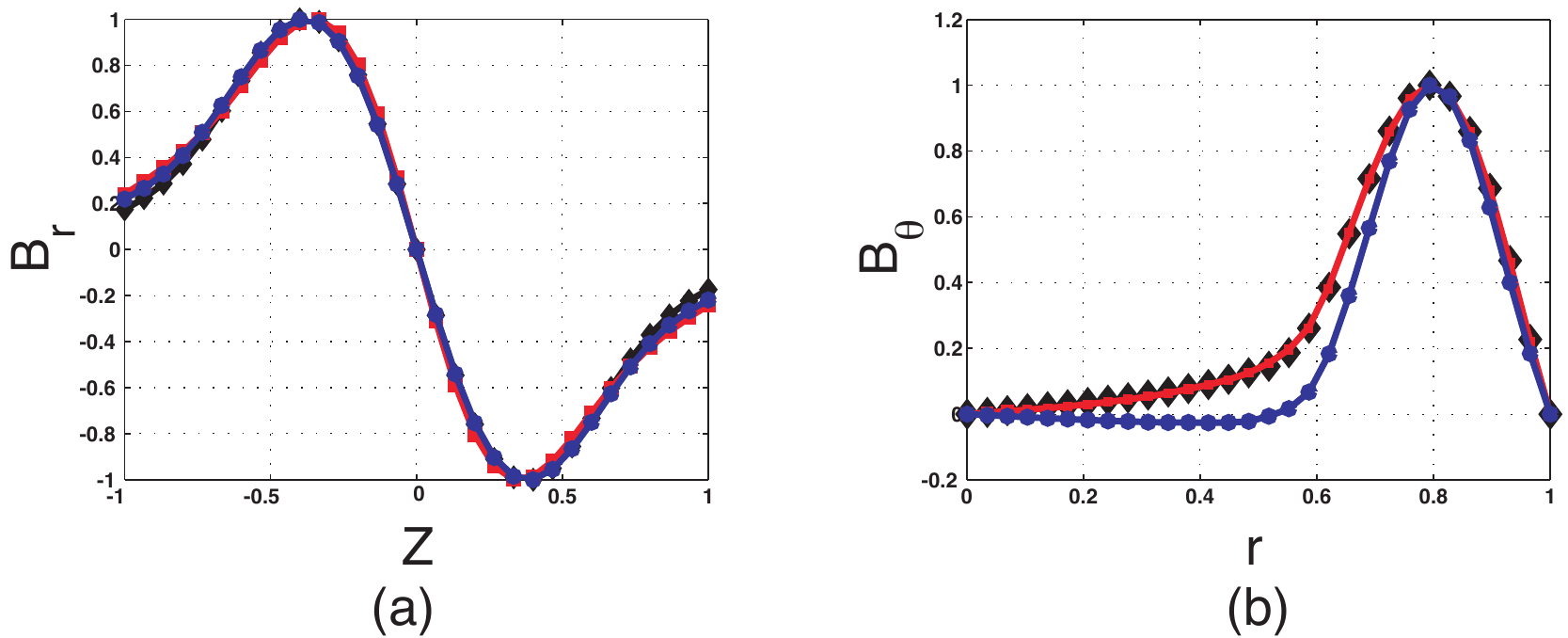}}
\caption{Dipolar mode: estimate of the error made doing the projection (a): axial profile of $B_r$ at $r=0.8$ from the $Oz$ axis. (b): radial profile of $B_\theta$
in the median plane. ($\blacksquare$) : Components of the applied field
($\blacklozenge$)  Components of $\langle\cL^2\bB^P_d\rangle$. ($\bullet$)
Components of  $\langle\cL^2\bB^T_d\rangle$.}
\label{fig:err_dipole}
\end{figure}

\newpage
\rightline{\bf Volk {\it et al.}, Figure 14}
\vspace*{10mm}
\begin{figure}[h]
\centerline{\includegraphics[width=15cm]{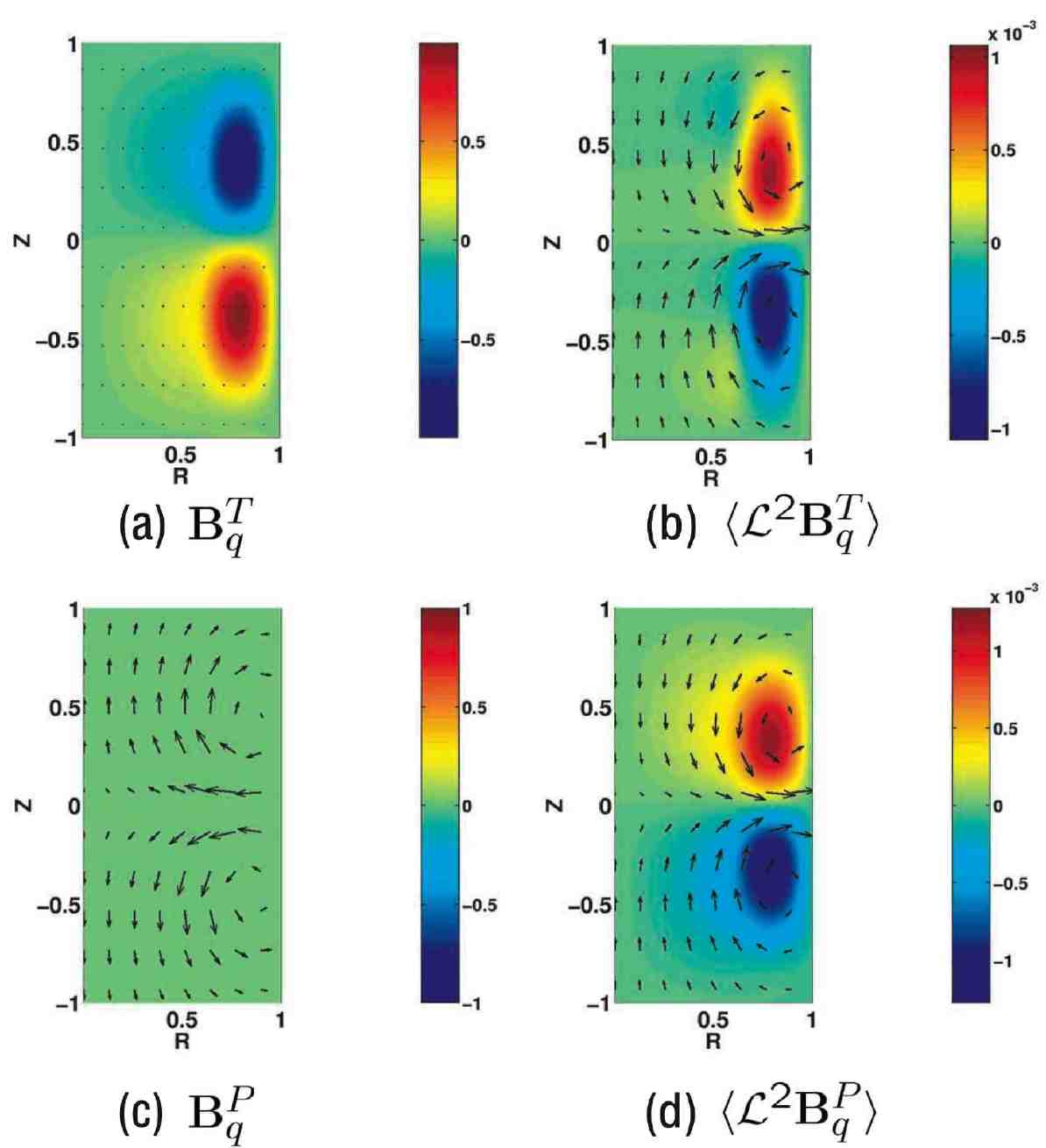}}
\caption{Quadrupolar mode: (a),(c) resp. toroidal and poloidal component of the quadrupole eigenvector. (b), (d), axisymmetric part of the fields obtained by applying $\cL^2$ to the vector fields (a) and (c).}
\label{fig:L_quadrupole}
\end{figure}

\newpage
\rightline{\bf Volk {\it et al.}, Figure 15}
\vspace*{10mm}
\begin{figure}[h]
\centerline{\includegraphics[width=16cm]{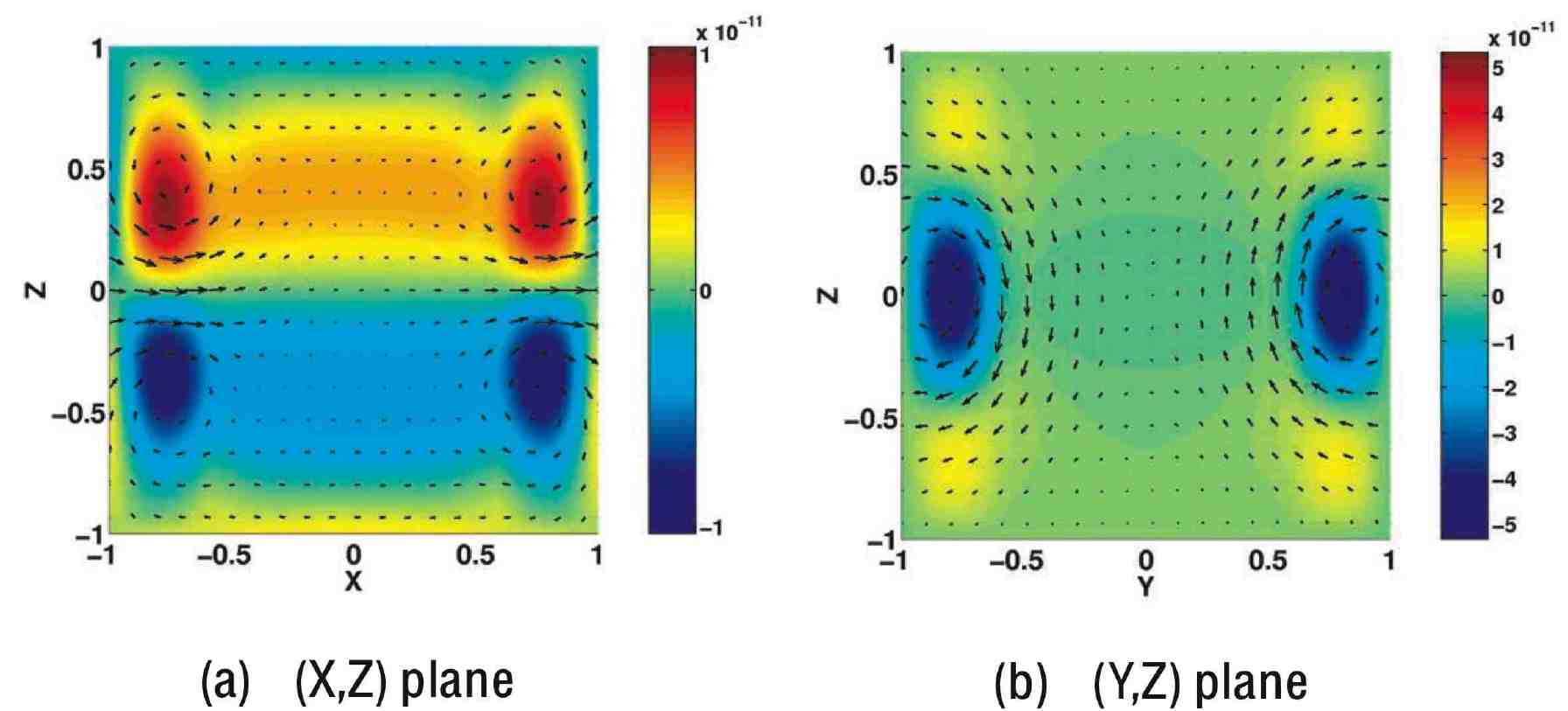}}
\caption{Transverse dipole mode: (a) ($Ox$,$Oz$) cut. (b)  ($Oy$,$Oz$) cut.}
\label{fig:transverse_mode}
\end{figure}

\newpage
\rightline{\bf Volk {\it et al.}, Figure 16}
\vspace*{10mm}
\begin{figure}[h]
\centerline{\includegraphics[width=17cm]{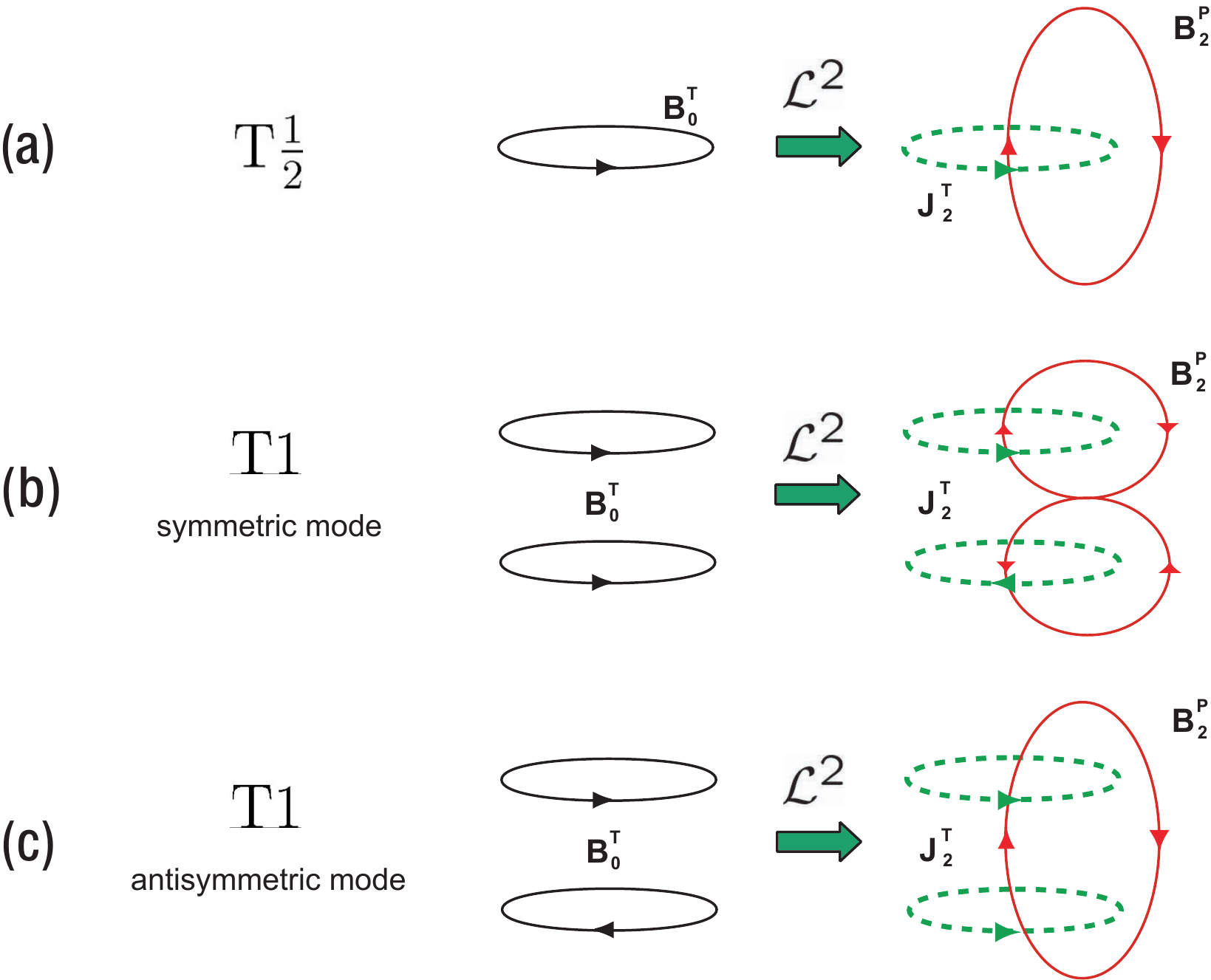}}
\caption{Comparison between the schematic induction mechanisms for the $T\frac{1}{2}$ and $T1$ flows, due to the $\alpha$ effect in the case of a toroidal applied field. The black line represents the applied field, the green dashed line the resulting order 2 current and the red line the resulting order 2 magnetic field. (a) $T\frac{1}{2}$ flow (b) $T1$ flow, with a symmetric applied field (c) $T1$ flow, with an antisymmetric applied field.}
\label{fig:modes_T1}
\end{figure}

\newpage
\rightline{\bf Volk {\it et al.}, Figure 17}
\vspace*{10mm}
\begin{figure}[h]
\centerline{\includegraphics[width=14cm]{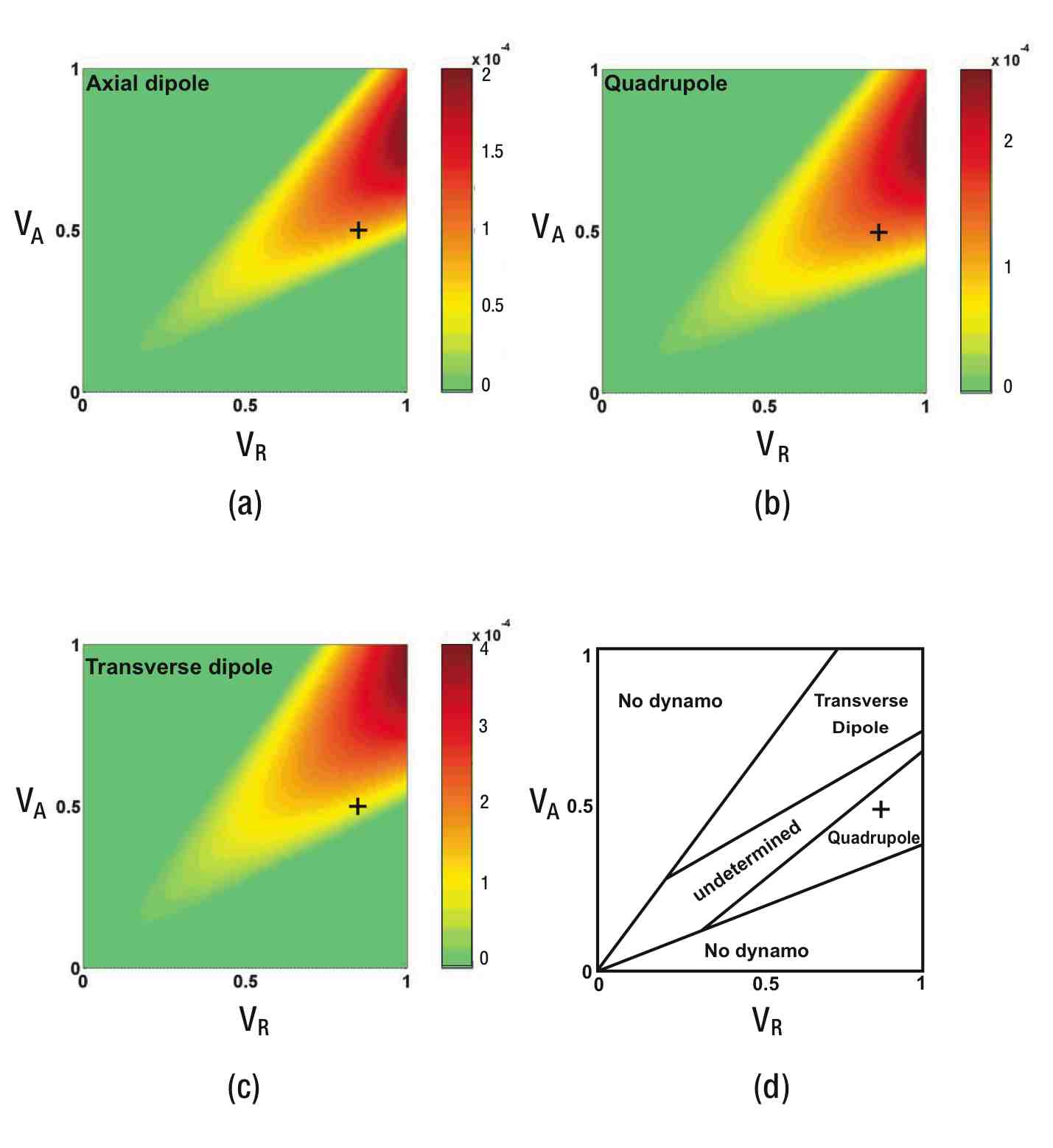}}
\caption{ $T\frac{1}{2}$ flow : evolution of the largest eigenvalue of the induction matrix in the ($V_R$,$V_A$) plane. (a) axial dipole. (b) axial quadrupole (c) transverse dipole (d) predominance diagram for the different possible modes. For simplicity, when the eigenvalue is negative we plotted a zero value. On each plot, the cross marks the parameter couple ($V_R$,$V_A$)=(0.83,0.53) of the particular flow studied in \ref{dynamo_Tdemi}.}
\label{fig:Tdemi_diag}
\end{figure}

\newpage
\rightline{\bf Volk {\it et al.}, Figure 18}
\vspace*{10mm}
\begin{figure}[h]
\centerline{\includegraphics[width=15cm]{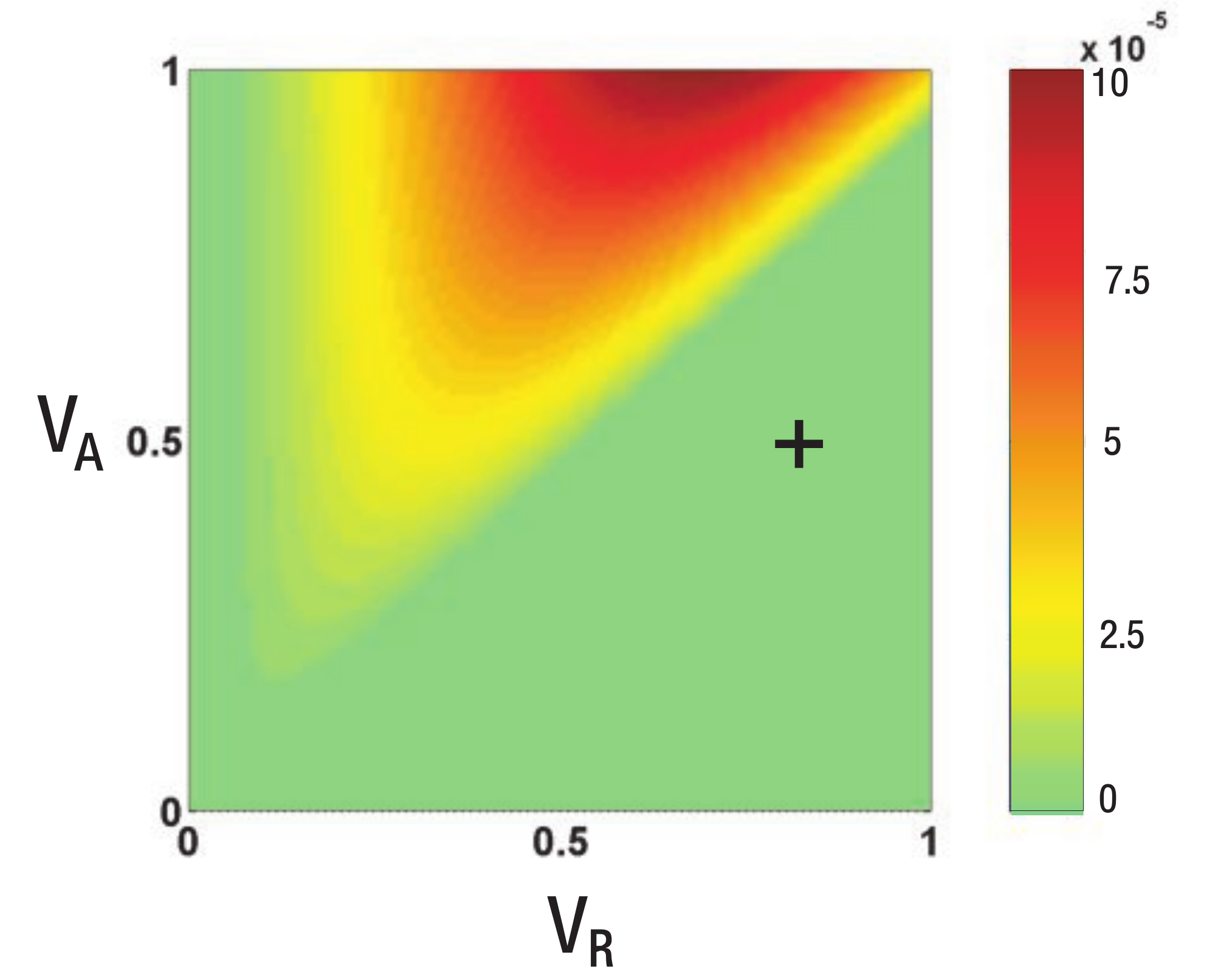}}
\caption{ $T1$ flow : evolution of the largest eigenvalue of the induction matrix in the ($V_R$,$V_A$) plane, for the axial quadrupole. For simplicity, when the eigenvalue is negative we plotted a zero value. The cross marks the parameter couple ($V_R$,$V_A$)=(0.83,0.53) of the particular flow studied in \ref{dynamo_T1}.}
\label{fig:T1_diag}
\end{figure}


\begin{thebibliography}{99}
\bibitem{CommentBz}
Since the outside medium is insulating and for a $z$-independent system, one can write $\bB = - \bnabla \psi$, with $\psi = z.f(r,\theta)+g(r,\theta)$, $f$ and $g$ being two arbitrary functions independent of the $z$ coordinate. Then $(B_r, B_\theta, B_z) = (-\partial_r g - z\partial_r f, - (1/r) \partial_\theta g - (z/r) \partial_\theta f, - f)$ so that the condition that $(B_r, B_\theta)$ be $z$-independent leads to $f \equiv 0$ (the field should vanish at infinity), and hence $B_z = 0$ at $r=1$.
\end{thebibliography}
\end{document}